\documentclass[3p,times]{elsarticle}
\usepackage{amssymb}
\usepackage{epsfig}
\usepackage{mathrsfs}
\usepackage{graphicx}
\usepackage{amsmath}
\usepackage{dcolumn}
\usepackage{bm}
\usepackage{longtable}
\usepackage{threeparttablex}
\usepackage{multirow}
\usepackage{booktabs}
\usepackage{color}
\usepackage{pdflscape}
\usepackage[figuresright]{rotating}
\usepackage[version=4]{mhchem}
\usepackage[labelfont=bf,singlelinecheck=false,font=footnotesize]{caption} 
\usepackage{hyperref}
\usepackage{array}
\newcolumntype{C}[1]{>{\centering\arraybackslash}p{#1}}
\biboptions{square,sort&compress} 
\newcommand{\beq}{\begin{equation}}
\newcommand{\eeq}{\end{equation}}
\newcommand{\ccp}{CC($P$)}
\newcommand{\eomccp}{EOMCC($P$)}
\newcommand{\ccpq}{CC($P$;$Q$)}
\newcommand{\AP}[2]{${#1}\:^{#2}\text{A}^{\prime}$}
\newcommand{\APP}[2]{${#1}\:^{#2}\text{A}^{\prime\prime}$}

\journal{Chemical Physics Letters}

\begin{document}
\begin{frontmatter}
\title{
Extension of the CIPSI-Driven \ccpq\ Approach to Excited Electronic States
}
\author[label1]{Swati S. Priyadarsini}
\author[label1]{Karthik Gururangan}
\author[label1,label2]{Piotr Piecuch\corref{cor1}}
\address[label1]{Department of Chemistry, Michigan State University, 
East Lansing, Michigan 48824, USA}
\address[label2]{Department of Physics and Astronomy, Michigan State University, East Lansing, Michigan 48824, USA}
\ead{piecuch@chemistry.msu.edu}
\cortext[cor1]{Corresponding author}

\begin{abstract}
We extend the CIPSI-driven \ccpq\ methodology
[K. Gururangan {\it et al.}, J. Chem. Phys. 155 (2021) 174114],
in which the leading higher--than--doubly excited determinants are identified using
the selected configuration interaction (CI) approach abbreviated as CIPSI, to excited
electronic states via the equation-of-motion (EOM) coupled-cluster (CC) formalism. By
examining vertical excitations in \ce{CH+} at equilibrium and stretched geometries,
adiabatic excitations in CH, and ground- and excited-state potential cuts of water,
we demonstrate that the CIPSI-driven \ccpq\ method converges
parent CC/EOMCC singles, doubles, and triples energetics from relatively inexpensive
Hamiltonian diagonalizations in CI spaces smaller than the corresponding triples manifolds.
\end{abstract}


\end{frontmatter}

%
\section{Introduction}
\label{sec1}
An accurate determination of excited electronic states, which lie at the heart of
spectroscopy and photochemistry, remains one of the biggest challenges of contemporary
quantum chemistry, especially when excited states dominated by two- or other
many-electron transitions and excited-state potential energy surfaces (PESs) along
bond-stretching coordinates are examined. Meeting this challenge requires the development
of reliable, yet practical, methods that can deliver high-quality and
well-balanced results for excitation as well as total ground- and excited-state energies
at manageable computational costs. In this work, we focus on a promising new approach
that can achieve these objectives within the equation-of-motion (EOM) extension
\cite{eomcc3}
of single-reference coupled-cluster (CC) theory \cite{cizek1,cizek4}.

In the EOMCC formalism and its closely related linear-response (LR)
\cite{monk}
CC and symmetry-adapted-cluster configuration interaction (CI) \cite{sacci4} counterparts,
the excited-state wave functions of an $N$-electron system are expressed as
$|\Psi_{\mu}\rangle = R_{\mu}|\Psi_{0}\rangle = R_{\mu}e^{T}|\Phi\rangle$, where
$|\Phi\rangle$ is the reference determinant
defining the Fermi vacuum 
and $T = \sum_{n=1}^{N}T_{n}$
and $R_{\mu} = \sum_{n=0}^{N}R_{\mu,n} = r_{\mu,0}{\bf 1} + \sum_{n=1}^{N}R_{\mu,n}$ are
the cluster and EOM excitation operators, respectively, with $T_{n}$ and $R_{\mu,n}$ representing their
$n$-body components and {\bf 1} denoting the unit operator. The basic EOMCC approach with singles
and doubles (EOMCCSD) \cite{eomcc3}, which is obtained by truncating $T$ and $R_{\mu}$
at their two-body components and characterized by relatively inexpensive
computational steps that scale with the system size $\mathscr{N}$ as $\mathscr{O}(\mathscr{N}^{6})$,
is useful in describing excited states dominated by one-electron transitions, but it fails
in more multireference (MR) situations, such as excited-state potentials along bond-stretching
coordinates and excited states with substantial double-excitation character
\cite{eomccsdt1,eomccsdt2,water,kgjspp_h2o_2025,creomcc-2015} (EOMCCSD may struggle
with singly excited states too \cite{creomcc-2015}). A conceptually
straightforward remedy to most of the problems encountered in EOMCCSD computations
is the next method in the EOMCC hierarchy, namely, the EOMCC approach
with a full treatment of singles, doubles, and triples (EOMCCSDT)
\cite{eomccsdt1,eomccsdt2,eomccsdt3}, in which $T$ and $R_{\mu}$ are truncated at $T_{3}$ and
$R_{\mu,3}$, respectively, but the improvements in the EOMCCSD results offered by EOMCCSDT
come at a very high price, as the iterative steps characterizing EOMCCSDT calculations scale as
$\mathscr{O}(\mathscr{N}^{8})$, rendering them prohibitively expensive for all but
very small molecules. It is, therefore, desirable to develop
approximations to full EOMCCSDT capable of recovering EOMCCSDT energetics at small fractions
of the computational effort. Over the years, a variety of EOMCC and LRCC schemes have been
proposed that use arguments originating from many-body perturbation theory (MBPT) to correct
EOMCCSD or LRCCSD excitation energies for the leading $T_{3}$ and $R_{\mu,3}$ correlations
in an iterative or noniterative fashion \cite{eomap3,ccsdr3a,eomap5,cc3_1}, but the resulting
approaches, while accurate for excited states dominated by one-electron transitions, are generally
insufficient for states with larger contributions from doubly excited configurations and for
excited-state potentials along bond-stretching coordinates. The more robust completely renormalized
(CR) triples corrections to EOMCCSD \cite{water,creomcc-2015,kkppeom,crccl_molphys,crccl_ijqc2,7hq},
such as CR-EOMCC(2,3) \cite{crccl_molphys,crccl_ijqc2,jspp-chemphys2012}
and its size-intensive $\delta$-CR-EOMCC(2,3) \cite{creomcc-2015,7hq} extension,
their analogs based on partitioning the similarity-transformed Hamiltonian \cite{eomccpt,eomccpt2},
and the active-space EOMCCSDt approach \cite{eomccsdt1,eomccsdt2}, can handle doubly
excited states and excited-state potentials, but they
fail to accurately approximate the parent EOMCCSDT energetics when
the coupling between the $T_n$ and $R_{\mu,n}$ amplitudes
with $n \leq 2$ and their higher-rank $T_{3}$ and $R_{\mu,3}$ counterparts becomes large
\cite{water,stochastic-ccpq-molphys-2020,kgjspp_h2o_2025}.

As shown in Refs.\  \cite{stochastic-ccpq-molphys-2020,kgjspp_h2o_2025}, problems encountered in
calculations using perturbative, CR-type, and active-space approximations to EOMCCSDT can be alleviated
by adopting the \ccpq\ framework introduced in Ref.\ \cite{jspp-chemphys2012}. The \ccpq\ formalism,
which applies to ground
\cite{jspp-chemphys2012,jspp-jcp2012,jspp-jctc2012,nbjspp-molphys2017,ccpq-be2-jpca-2018,ccpq-mg2-mp-2019,%
stochastic-ccpq-prl-2017,stochastic-ccpq-jcp-2021,cipsi-ccpq-2021,arnab-stgap-2022,adaptiveccpq2023,cbd_cipsi_ccpq}
and excited \cite{jspp-chemphys2012,stochastic-ccpq-molphys-2020,kgjspp_h2o_2025} states,
generalizes the biorthogonal moment expansions that led to the aforementioned
CR-EOMCC(2,3) and $\delta$-CR-EOMCC(2,3) approaches and their ground-state CR-CC(2,3) counterpart \cite{crccl_jcp},
to unconventional truncations in the $T$ and $R_{\mu}$ operators.
By incorporating the leading contributions to the $T_{n}$ and $R_{\mu,n}$ components with $n>2$ into
the CC/EOMCC iterations, {\it i.e.}, by relaxing the lower-rank $T_1$, $T_2$, $R_{\mu,1}$, and $R_{\mu,2}$
amplitudes in the presence of their higher-rank counterparts, and capturing the remaining correlations
of interest
using suitably defined energy corrections, \ccpq\ calculations can accurately approximate
high-level CC/EOMCC energetics, such as those of CCSDT \cite{ccfullt,ccfullt2} and EOMCCSDT, with substantially
reduced computational costs, even when $T_{3}$ and $R_{\mu,3}$ amplitudes and MR effects become larger
\cite{jspp-chemphys2012,jspp-jcp2012,jspp-jctc2012,nbjspp-molphys2017,ccpq-be2-jpca-2018,ccpq-mg2-mp-2019,%
stochastic-ccpq-prl-2017,stochastic-ccpq-molphys-2020,stochastic-ccpq-jcp-2021,cipsi-ccpq-2021,arnab-stgap-2022,%
adaptiveccpq2023,cbd_cipsi_ccpq,kgjspp_h2o_2025}. The \ccpq\ approaches targeting CCSDT developed
to date include the active-orbital-based CC(t;3) method \cite{jspp-chemphys2012,jspp-jcp2012,jspp-jctc2012,%
nbjspp-molphys2017,ccpq-be2-jpca-2018,ccpq-mg2-mp-2019} and its more black-box semi-stochastic
\cite{stochastic-ccpq-prl-2017,stochastic-ccpq-molphys-2020,stochastic-ccpq-jcp-2021,arnab-stgap-2022},
adaptive \cite{adaptiveccpq2023}, and selected-CI-driven \cite{cipsi-ccpq-2021,cbd_cipsi_ccpq} counterparts.
The semi-stochastic, adaptive, and active-orbital-based \ccpq\ methods targeting EOMCCSDT have been implemented
too \cite{stochastic-ccpq-molphys-2020,kgjspp_h2o_2025}, but the analogous excited-state extension of the
\ccpq\ approach using the selected CI algorithm abbreviated as CIPSI \cite{sci_3,cipsi_1,cipsi_2}, which
showed considerable promise in ground-state applications \cite{cipsi-ccpq-2021,cbd_cipsi_ccpq}, has not yet
been developed. We address this gap in the present work. To test the efficiency of the excited-state CIPSI-driven
\ccpq\ method, which uses relatively small Hamiltonian diagonalizations to identify the dominant triply
excited determinants for inclusion in the iterative steps of the \ccpq\ procedure, in converging EOMCCSDT
energetics, we apply it to vertical excitations in the equilibrium and stretched \ce{CH+} ion, adiabatic
excitations in the CH radical, and ground- and 11 excited-state PES cuts of the water molecule along the
O--H bond breaking coordinate.
\section{Theory}
\label{sec2}
The \ccpq\ calculations for ground and excited electronic states employ a two-step workflow.
In the first step --
abbreviated as \ccp\ for the ground ($\mu = 0$) state and \eomccp\ for excited ($\mu > 0$) states --
we solve the CC/EOMCC equations in a subspace of the $N$-electron Hilbert space, designated as
$\mathscr{H}^{(P)}$ and referred to as the $P$ space, which consists of the excited determinants
$|\Phi_{K}\rangle = E_{K}|\Phi\rangle$ that, together with the reference determinant $|\Phi\rangle$,
dominate the ground- and excited-state wave functions $|\Psi_{\mu}\rangle$ of interest ($E_{K}$ denotes an
elementary particle--hole excitation operator generating $|\Phi_{K}\rangle$ from $|\Phi\rangle$). Thus,
we start by solving for the amplitudes $t_{K}$ that define the $P$-space cluster operator
\beq
T^{(P)} = \sum_{|\Phi_{K}\rangle \in \mathscr{H}^{(P)}} t_{K} E_{K}
\label{eq1}
\eeq
associated with the \ccp\ ground state $|\Psi_{0}^{(P)}\rangle = e^{T^{(P)}}|\Phi\rangle$
and energy $E_{0}^{(P)} = \langle\Phi|\overline{H}^{(P)}|\Phi\rangle$, where
$\overline{H}^{(P)} = e^{-T^{(P)}}He^{T^{(P)}}$ is the similarity-transformed Hamiltonian, and
diagonalizing $\overline{H}^{(P)}$ in the $P$ space to determine the \eomccp\ excited-state energies
$E_{\mu}^{(P)}$ and the corresponding particle--hole and hole--particle excitation and deexcitation operators,
\beq
R_{\mu}^{(P)} = r_{\mu,0}{\bf 1} + \sum_{|\Phi_{K}\rangle \in \mathscr{H}^{(P)}}r_{\mu,K}E_{K}
\label{eq2}
\eeq
and
\beq
L_{\mu}^{(P)} = \delta_{\mu,0}{\bf 1} + \sum_{|\Phi_{K}\rangle \in \mathscr{H}^{(P)}}l_{\mu,K}(E_{K})^{\dagger},
\label{eq3}
\eeq
respectively, where the $r_{\mu,K}$ and $l_{\mu,K}$ coefficients define the \eomccp\ ket states
$|\Psi_{\mu}^{(P)}\rangle = R_{\mu}^{(P)}e^{T^{(P)}}|\Phi\rangle$ and the \ccp\ ($\mu=0$) or
\eomccp\ ($\mu>0$) bra states
$\langle\tilde{\Psi}_{\mu}^{(P)}| = \langle\Phi|L_{\mu}^{(P)}e^{-T^{(P)}}$ satisfying the
biorthonormality condition $\langle\tilde{\Psi}_{\mu}^{(P)}|\Psi_{\nu}^{(P)}\rangle = \delta_{\mu,\nu}$
($\delta_{\mu,\nu}$ is the usual Kronecker delta). Once
$T^{(P)}$, $L_{0}^{(P)}$, and $E_{0}^{(P)}$ and, in the case of excited states, $R_{\mu}^{(P)}$, $L_{\mu}^{(P)}$,
and $E_{\mu}^{(P)}$ ($\mu > 0$) are obtained, we proceed to the second step of the \ccpq\ procedure, in which
the \ccp\ and \eomccp\ energies $E_{\mu}^{(P)}$ are corrected for the remaining correlation effects using
the state-specific noniterative corrections
\beq
\delta_{\mu}(P;\!Q) = \sum_{|\Phi_{K}\rangle \in \mathscr{H}^{(Q)}}\ell_{\mu,K}(P)\:\mathfrak{M}_{\mu,K}(P) ,
\label{eq4}
\eeq
calculated with the help of another subspace of the $N$-electron Hilbert space, referred to as the $Q$ space,
denoted as $\mathscr{H}^{(Q)}$ [$\mathscr{H}^{(Q)} \subseteq (\mathscr{H}^{(0)} \oplus \mathscr{H}^{(P)})^{\perp}$,
where $\mathscr{H}^{(0)} = \mathrm{span}\{|\Phi\rangle$\}]. 
The quantities $\mathfrak{M}_{\mu,K}(P)$ entering Eq.\ (\ref{eq4}), defined as
$\mathfrak{M}_{0,K}(P) = \langle \Phi_K | \overline{H}^{(P)}|\Phi\rangle$ for the ground state and
$\mathfrak{M}_{\mu,K}(P) = \langle \Phi_K | \overline{H}^{(P)} R_{\mu}^{(P)} |\Phi\rangle$ for excited states,
are the generalized moments of the CC($P$) ($\mu = 0$) and EOMCC($P$) ($\mu > 0$) equations representing
projections of these equations on the $Q$-space determinants $|\Phi_K\rangle \in \mathscr{H}^{(Q)}$.
The coefficients $\ell_{\mu,K}(P)$ that multiply moments $\mathfrak{M}_{\mu,K}(P)$ in Eq.\ (\ref{eq4})
are calculated as $\ell_{\mu,K}(P) = \langle \Phi| L_{\mu}^{(P)}\overline{H}^{(P)} |\Phi_K \rangle / D_{\mu,K}^{(P)}$,
where $D_{\mu,K}^{(P)} = E_{\mu}^{(P)} - \langle \Phi_K | \overline{H}^{(P)} | \Phi_K \rangle$ is the
Epstein--Nesbet-style denominator. The final \ccpq\ energies are determined using the formula
\beq
E_{\mu}^{(P + Q)} = E_{\mu}^{(P)} + \delta_{\mu}(P;\!Q).
\label{eq5}
\eeq
All that is needed
now
is a reliable procedure for designing the $P$ and $Q$ spaces that would allow us to accurately and
efficiently capture the many-electron correlation effects characterizing the ground
and excited states of interest. In this study, where our objective is to accurately approximate CCSDT/EOMCCSDT
energetics by extending the CIPSI-driven \ccpq\ methodology
of Refs.\ \cite{cipsi-ccpq-2021,cbd_cipsi_ccpq}
to excited states, the $P$ spaces used in the iterative \ccp\ and \eomccp\ steps are spanned by
all singly and doubly excited determinants and the leading triply excited determinants identified with the
help of
the CIPSI algorithm. The complementary
$Q$ spaces, needed to evaluate the \ccpq\ corrections $\delta_{\mu}(P;\!Q)$,
are defined
as the remaining triply excited determinants not captured by CIPSI, such that the union of the $P$ and $Q$ spaces
always consists of all singly, doubly, and triply excited determinants.

We recall that the CIPSI method, proposed in Ref.\ \cite{sci_3} and subsequently developed in Refs.\
\cite{cipsi_1,cipsi_2}, constructs an approximation to full CI through a sequence of
Hamiltonian diagonalizations in increasingly large, iteratively defined, subspaces of the many-electron Hilbert
space, denoted as $\mathcal{V}_{\rm int}^{(k)}$, where $k = 0, 1, 2, \dots$ enumerates consecutive CIPSI iterations.
After defining the initial subspace $\mathcal{V}_{\rm int}^{(0)}$, which in the calculations carried
out in this study consisted of the restricted Hartree--Fock (RHF) or restricted open-shell Hartree--Fock (ROHF)
determinant for the ground state and the leading configuration-state functions (CSFs) for the lowest-energy states
belonging to irreducible representations (irreps) other than that of the ground state, each subsequent subspace
$\mathcal{V}_{\rm int}^{(k+1)}$ with $k\geq0$ is obtained by augmenting its $\mathcal{V}_{\rm int}^{(k)}$
predecessor with a subset of the leading singly and doubly excited determinants generated from it identified
with the help of MRMBPT, adding them, one by one, to $\mathcal{V}_{\rm int}^{(k)}$ until the dimension of
$\mathcal{V}_{\rm int}^{(k+1)}$ exceeds that of $\mathcal{V}_{\rm int}^{(k)}$ by a user-defined factor
$f > 1$ (the  number of determinants
in $\mathcal{V}_{\rm int}^{(k+1)}$ is usually somewhat
larger than $f$ times the dimension of $\mathcal{V}_{\rm int}^{(k)}$, since additional determinants may have to be
incorporated in $\mathcal{V}_{\rm int}^{(k+1)}$ to ensure that the resulting wave functions are eigenfunctions of the
total spin operators $S^{2}$ and $S_{z}$).
The CIPSI diagonalization sequence is terminated either when the
second-order MRMBPT corrections, used
to correct the raw CIPSI energies obtained in Hamiltonian diagonalizations,
fall below a user-specified threshold $\eta$, or when the
number of determinants in $\mathcal{V}_{\rm int}^{(k+1)}$ exceeds
an input parameter $N_{\rm det(in)}$.
In this work, where we examine the convergence of the CIPSI-driven \ccpq\ energies toward CCSDT/EOMCCSDT as a function
of $N_{\rm det(in)}$, we follow Refs.\ \cite{cipsi-ccpq-2021,cbd_cipsi_ccpq} and adopt the latter termination
criterion. We stop when the number of determinants in the final wave function
$|\Psi^{\rm(CIPSI)}\rangle$, $N_{\rm det(out)}$, of a CIPSI run providing the list of triply excited determinants
for inclusion in the $P$ space satisfies $N_{\rm det(out)} \geq N_{\rm det(in)}$.

With the above information in mind, the key steps of the CIPSI-based \ccpq\ algorithm for ground and excited
states developed
in this work, designed to recover the CCSDT/EOMCCSDT-quality energetics,
are as follows:
\begin{enumerate}
\item Choose a wave function termination parameter $N_{\rm det(in)}$ and run a CIPSI calculation starting from
a suitably chosen zeroth-order subspace $\mathcal{V}_{\rm int}^{(0)}$ (spanned by the RHF or ROHF determinant)
to obtain the ground $|\Psi^{\rm(CIPSI)}\rangle$ state. If there is interest in electronic states
belonging to irreps other than that of the ground state, execute the analogous CIPSI
runs for lowest-energy states of these other irreps as well.
\item Extract a list or, if
states belonging to multiple irreps are targeted,
lists of triply excited determinants from the $|\Psi^{\rm(CIPSI)}\rangle$ wave function(s) obtained in Step 1 to
determine the $P$ space(s) for the \ccp\ and \eomccp\ calculations. For the ground
state and excited states belonging to the same irrep as the ground state, the $P$ space entering the \ccp/\eomccp\
calculations consists of all singly and doubly excited determinants and the subset of triply excited determinants
captured by the CIPSI ground state. For the excited states belonging to other irreps, the $P$ space defining the
\ccp\ problem is the same as that used for the ground state, but the lists of triples entering the \eomccp\
diagonalizations are extracted from the CIPSI computations for the lowest-energy states of those other irreps.
\item Solve the \ccp\ and, if excited states are targeted, \eomccp\ equations in the $P$ space or spaces obtained in
Step 2. For the ground state and excited states belonging to the same irrep as the ground state, we define
$T^{(P)} = T_{\rm 1} + T_{\rm 2} + T_{\rm 3}^{\rm (CIPSI)}$,
$R_{\mu}^{(P)} = r_{\mu,0}{\bf 1} + R_{\mu,\rm 1} + R_{\mu,\rm 2} + R_{\mu,\rm 3}^{\rm (CIPSI)}$, and
$L_{\mu}^{(P)} = \delta_{\mu,0}{\bf 1} + L_{\mu,\rm 1} + L_{\mu,\rm 2} + L_{\mu,\rm 3}^{\rm (CIPSI)}$,
where the list of triples in $T_{\rm 3}^{\rm (CIPSI)}$, $R_{\mu,\rm 3}^{\rm (CIPSI)}$, and $L_{\mu,\rm 3}^{\rm (CIPSI)}$
is extracted from the CIPSI ground state. For the excited states belonging to other irreps, we construct
the similarity-transformed Hamiltonian $\overline{H}^{(P)}$, to be diagonalized in the \eomccp\ calculations,
in the same way as in the ground-state computations, but
$R_{\mu,\rm 3}^{\rm (CIPSI)}$ and $L_{\mu,\rm 3}^{\rm (CIPSI)}$ are defined using the lists of triples
contained in the $|\Psi^{\rm(CIPSI)}\rangle$ wave functions that are the lowest-energy states of those irreps.
\item 
Determine corrections $\delta_{\mu}$($P$;$Q$), Eq.\ (\ref{eq4}), describing the remaining $T_{3}$ and $R_{\mu,\rm 3}$
correlations not captured by the CIPSI-driven \ccp\ and \eomccp\ calculations, by defining the respective
$Q$ space(s) as those triply excited determinants that are outside the $P$ space(s) considered
in Steps 2 and 3. Add the resulting corrections $\delta_{\mu}$($P$;$Q$) to the \ccp/\eomccp\ energies
$E_{\mu}^{(P)}$ to obtain the final \ccpq\ energies $E_{\mu}^{(P+Q)}$, Eq.\ (\ref{eq5}).
\item To check convergence, repeat Steps 1--4 for some larger values of $N_{\rm det(in)}$. The CIPSI-driven \ccpq\
calculations can be regarded as converged if the differences between consecutive $E_{\mu}^{(P+Q)}$ energies fall
below a user-defined threshold. One can also consider stopping the calculations
if the fraction(s) of triples contained in the final CIPSI state(s)
$|\Psi^{\rm(CIPSI)}\rangle$ is (are) sufficiently large to produce the desired accuracy level with Eq.\ (\ref{eq5}).
\end{enumerate}
%
The above points summarize a specific implementation of the CIPSI-driven \ccpq\ approach pursued in the
present study, in which the $P$ spaces adopted in the \eomccp\ computations are based on the lowest state in
each irrep of interest, but other possibilities exist. We could, for example, consider an alternative
formulation, in which one runs CIPSI for multiple electronic states in each symmetry block [{\it i.e.},
all the states targeted by \ccpq] and constructs lists of higher--than--doubly excited determinants for
inclusion in the corresponding $P$ spaces specific to the targeted individual states or their groups.
Work on developing such an algorithm is currently underway in our group.

Similarly to the ground-state CIPSI-driven \ccpq\ approach of Refs.\ \cite{cipsi-ccpq-2021,cbd_cipsi_ccpq},
the above algorithm offers tremendous savings in computational effort relative to CCSDT/EOMCCSDT, which
originate from three factors. First, as shown in the next section, the underlying CIPSI runs, needed to
accurately approximate the parent CCSDT/EOMCCSDT energetics using the \ccpq\ method summarized above,
rely on small $N_{\rm det(in)}$ values
resulting in compact CI spaces that are much smaller than the
numbers of all triples employed in CCSDT/EOMCCSDT.
Second, the \ccp\ and \eomccp\ calculations using tiny fractions of triples in the corresponding $P$ spaces,
such as those seen in the numerical examples discussed in this work, are
one or more orders of magnitude faster than their CCSDT and EOMCCSDT counterparts.
Third, the computational costs associated with the determination of the noniterative corrections
$\delta_{\mu}$($P$;$Q$), which are similar -- per state -- to those characterizing
the triples corrections of CCSD(T), CR-CC(2,3), or CR-EOMCC(2,3), are smaller than those of a
single iteration of CCSDT/EOMCCSDT. The key elements of our algorithm used to implement the \ccp\ equations,
designed to efficiently handle small but generally spotty subsets of triply excited determinants identified
by CIPSI, along with illustrative computational timings characterizing the ground-state \ccpq\ calculations,
are described in the Appendix of Ref.\ \cite{cbd_cipsi_ccpq}. We adopt a similar strategy when solving
the \eomccp\ equations. A detailed description of our efficient implementation of the
\ccp\ and \eomccp\ equations and the \ccpq\ corrections targeting CCSDT/EOMCCSDT will be presented in a
future publication.
\section{Numerical examples}
\label{sec3}
\subsection{Computational details}
\label{sec3.1}
To assess the ability of the excited-state extension of the CIPSI-driven \ccpq\ approach
to converge EOMCCSDT energetics, we applied it to three molecular problems:
(i) the frequently examined vertical excitations in the \ce{CH+} ion, as described by the [5s3p1d/3s1p] basis
set of Ref.\ \cite{chplusbasis},
for which full EOMCCSDT is virtually exact \cite{eomccsdt1,eomccsdt2}, that were used to test
the semi-stochastic variant of \ccpq\ in Ref.\ \cite{stochastic-ccpq-molphys-2020},
(ii) the adiabatic excitations in the CH radical, as described by the aug-cc-pVDZ basis \cite{ccpvnz,augccpvnz},
used to test the semi-stochastic \ccpq\ method in Ref.\ \cite{stochastic-ccpq-molphys-2020}
as well, which, in analogy to \ce{CH+}, has low-lying excited states dominated by two-electron
transitions requiring the EOMCCSDT theory level to obtain a reliable description \cite{hirata1,crccl_ijqc2}, and
(iii) the ground- and excited-state PESs of the water molecule, as described by the TZ basis set of Ref.\
\cite{nmapp3}, corresponding to the $\ce{H2O} \rightarrow \ce{H} + \ce{OH}$ dissociation,
which, as demonstrated in Ref.\ \cite{kgjspp_h2o_2025}, require the
CCSDT/EOMCCSDT-level treatment to accurately approximate the full CI data and which were used to test the
CC(t;3) and adaptive \ccpq\ algorithms in Ref.\ \cite{kgjspp_h2o_2025}. Our calculations for the
\ce{CH+} and CH systems were performed using the largest Abelian subgroup of $C_{\infty v}$, namely, $C_{2v}$.
For the water molecule, where we were stretching one of the two O--H bonds, we used the relevant $C_{s}$ symmetry.
The \ccp/\eomccp, \ccpq\, and CCSDT/EOMCCSDT computations for \ce{CH+} and water relied on the RHF references,
whereas the reference determinants used for CH were obtained using ROHF. Consistent with our prior
semi-stochastic, adaptive, and active-orbital-based \ccpq\ studies of \ce{CH+}, CH, and water
\cite{kgjspp_h2o_2025,stochastic-ccpq-molphys-2020}, in the post-ROHF calculations for CH and
post-RHF computations for \ce{H2O}, the lowest core orbitals correlating with 1s shells of
the carbon and oxygen atoms were frozen. In the case of \ce{CH+}, we correlated all electrons.
As in Refs.\ \cite{cipsi-ccpq-2021,cbd_cipsi_ccpq}, the \ccp, \eomccp, and \ccpq\ results
discussed in the next three subsections,
along with their CCSDT and EOMCCSDT parents, were obtained using our open-source CCpy package available on
GitHub \cite{CCpy-GitHub}, which is interfaced with the RHF, ROHF, and integral-transformation routines in
GAMESS \cite{gamess2020}, whereas the preceding CIPSI runs were executed with Quantum Package 2.0 \cite{cipsi_2}.
In running CIPSI, we used the $N_{\rm det(in)}$ values in the 1--20,000 range, so that the Hamiltonian
diagonalization spaces generated by CIPSI were smaller than the numbers of triples used by CCSDT/EOMCCSDT,
and we set the subspace enlargement factor $f$ to its default value of 2 and the stopping parameter $\eta$
to $10^{-6}$ hartree. With these choices of $\eta$ and $f$, the CIPSI diagonalization sequences
were terminated when the numbers of determinants in the $|\Psi^{\rm (CIPSI)}\rangle$
wave functions used to identify the subsets of triply excited determinants for
inclusion in the $P$ spaces adopted in our \ccp, \eomccp, and \ccpq\ calculations exceeded $N_{\rm det(in)}$
and the $N_{\rm det(out)}$ values characterizing the $|\Psi^{\rm (CIPSI)}\rangle$ states
were always between $N_{\rm det(in)}$ and $2N_{\rm det(in)}$. The convergence threshold used in the
\ccp/\eomccp\ and parent CCSDT/EOMCCSDT calculations was set to $10^{-7}$ hartree.
%
\subsection{\ce{CH+}}
\label{sec3.2}
Our first example is the \ce{CH+} ion at the equilibrium ($R = R_{\rm e} = 2.13713$ bohr;
Table\ \ref{table1}) and stretched ($R = 2R_{\rm e}$; Table\ \ref{table2}) geometries. For each C--H distance
$R$, we examined vertical excitations corresponding to transitions from the ground state ($1\,^{1}\Sigma^{+}$)
to the three lowest excited states of $^{1}\Sigma^{+}$ symmetry,
two lowest $^{1}\Pi$ states,
and two lowest $^{1}\Delta$ states.
In the case of the $n \, ^{1}\Sigma^{+}$ ($n = 1\mbox{--}4$) states,
the lists of triples entering the three-body components of the $T^{(P)}$, $R_{\mu}^{(P)}$, and $L_{\mu}^{(P)}$
operators were extracted from the ground-state CIPSI runs initiated with the one-dimensional subspace
$\mathcal{V}_{\rm int}^{(0)}$ spanned by the RHF determinant. For the $n \, ^{1}\Pi$ and $n \, ^{1}\Delta$
($n = 1,2$) states, the three-body component of the cluster operator $T^{(P)}$, needed to construct
the similarity-transformed Hamiltonian $\overline{H}^{(P)}$, was obtained in the same way as for the
$^{1}\Sigma^{+}$ states, but the lists of triples defining $R_{\mu,3}^{\rm(CIPSI)}$ and
$L_{\mu,3}^{\rm(CIPSI)}$ in the subsequent \eomccp\ diagonalizations were provided by the CIPSI runs
targeting the lowest-energy states relevant to $^{1}\Pi$ and $^{1}\Delta$ symmetries, namely,
the $^{1}{\rm B}_{1}(C_{2v})$ component of the $1\,^{1}\Pi$ state for the $n \, ^{1}\Pi$ states
and the $^{1}{\rm A}_{2}(C_{2v})$ component of the $1\,^{1}\Delta$ state for the $n \, ^{1}\Delta$ states
(initiated with the appropriate $3\sigma\rightarrow1\pi$ and $3\sigma^{2}\rightarrow1\pi^{2}$ CSFs in
the respective $\mathcal{V}_{\rm int}^{(0)}$ subspaces).

To highlight the importance of enriching the $P$ spaces used in the \ccp\ and \eomccp\ calculations
with the leading triply excited determinants prior to evaluating the \ccpq\ corrections, we first
discuss the results obtained using $N_{\rm det(in)} = 1$, where the $P$ spaces contain only singles
and doubles. The \eomccp/$N_{\rm det(in)} = 1$ approach is equivalent to EOMCCSD. Thus, it is not
surprising that the \eomccp/$N_{\rm det(in)} = 1$ results for the $2\,^{1}\Sigma^{+}$, $2\,^{1}\Pi$,
$1 \, ^{1}\Delta$, and $2 \, ^{1}\Delta$ states at $R = R_{\rm e}$ and for all seven excited states of
\ce{CH+} at $R = 2R_{\rm e}$ considered in this study, which are dominated by two-electron
transitions \cite{eomccsdt1,eomccsdt2} (with the $2\,^{1}\Delta$ state at $R = 2R_{\rm e}$ also exhibiting
substantial triple excitation character), are poor, producing errors relative to EOMCCSDT that are about
20, 12, 34, 35, and 14--144 millihartree, respectively. The \eomccp/$N_{\rm det(in)} = 1$ calculations for
the $3\,^{1}\Sigma^{+}$, $4\,^{1}\Sigma^{+}$, and $1\,^{1}\Pi$ states at $R = R_{\rm e}$, which are dominated
by single excitations, are more accurate, but the 3--6 millihartree errors relative to EOMCCSDT remain.
The \ccpq/$N_{\rm det(in)} = 1$ corrections -- equivalent to the triples corrections of CR-CC(2,3)/
CR-EOMCC(2,3) -- are helpful, reducing the errors relative to EOMCCSDT
in most cases to 1--3 millihartree, but they are ineffective for the $4\,^{1}\Sigma^{+}$ and $2\,^{1}\Delta$
states at $R = 2R_{\rm e}$. For these two states, the energies obtained in the \ccpq/$N_{\rm det(in)} = 1$ calculations
deviate from their EOMCCSDT counterparts by about 13 and 63 millihartree, respectively, which is telling
us that to achieve an accurate description, the one- and two-body components of the cluster and EOM excitation
and deexcitation operators must be relaxed in the presence of their three-body counterparts before evaluating
the triples corrections. The CIPSI-driven \ccpq\ methodology enables this by increasing the $N_{\rm det(in)}$
parameter and introducing the leading triply excited determinants into the $P$ spaces employed in the \ccp\ and
\eomccp\ computations.

Indeed, as shown in Tables\ \ref{table1} and \ref{table2}, with as little as 1,178--1,792 $S_{z} = 0$ determinants
of the ${\rm A}_{1}(C_{2v})$, ${\rm B}_{1}(C_{2v})$, and ${\rm A}_{2}(C_{2v})$ symmetries in the terminal wave
functions $|\Psi^{\rm (CIPSI)}\rangle$ corresponding to the lowest-energy $^{1}\Sigma^{+}$, $^{1}\Pi$, and
$^{1}\Delta$ states of \ce{CH+} generated by the inexpensive CIPSI runs using $N_{\rm det(in)} = 1,000$,
which capture tiny fractions, on the order of 0.7--3.7\%, of the $31,912$ ${\rm A}_{1}(C_{2v})$-symmetric, $27,180$
${\rm B}_{1}(C_{2v})$-symmetric, and $22,012$ ${\rm A}_{2}(C_{2v})$-symmetric $S_{z} = 0$ triples used by
CCSDT/EOMCCSDT, the differences between the energies of the $n \, ^{1}\Sigma^{+}$ ($n = 2\mbox{--}4$),
$n \, ^{1}\Pi$ ($n = 1,2$), and $n \, ^{1}\Delta$ ($n = 1,2$) excited states of
\ce{CH+} at $R = R_{\rm e}$ obtained with the CIPSI-driven \ccpq\ approach and their EOMCCSDT counterparts do not
exceed 0.808 millihartree, being usually much smaller.
At $R = 2R_{\rm e}$, they do not exceed 1.282 millihartree, again being usually smaller.
In particular, the large, 12.657 and 63.405 millihartree,
errors relative to EOMCCSDT resulting from the \ccpq/$N_{\rm det(in)} = 1$ or
CR-EOMCC(2,3) calculations for the $4\,^{1}\Sigma^{+}$ and $2\,^{1}\Delta$ states of \ce{CH+} at $R = 2R_{\rm e}$
reduce to less than 1 millihartree
when the \ccpq/$N_{\rm det(in)} = 1,000$ method is employed. With the relatively small additional
effort corresponding to $N_{\rm det(in)} = 5,000$, which results in $5,392$--$9,446$ $S_{z} = 0$ determinants of the
${\rm A}_{1}(C_{2v})$, ${\rm B}_{1}(C_{2v})$, and ${\rm A}_{2}(C_{2v})$ symmetries in the final
Hamiltonian diagonalization spaces adopted in the CIPSI-driven
\ccpq\ runs, and only 5.7--16.8\% of all triples in the underlying $P$ spaces, the differences between
the energies of the $n \, ^{1}\Sigma^{+}$ ($n = 2\mbox{--}4$), $n \, ^{1}\Pi$ ($n = 1,2$),
and $n\, ^{1}\Delta$ ($n = 1,2$) excited
states of \ce{CH+} obtained in the CIPSI-driven \ccpq\ calculations and their EOMCCSDT parents reduce
even further, to 0.005--0.207
millihartree at $R = R_{\rm e}$ and 0.001--0.109 millihartree at $R = 2 R_{\rm e}$ (0.109 and 0.011 millihartree for the
$4\,^{1}\Sigma^{+}$ and $2\,^{1}\Delta$ states at $R = 2R_{\rm e}$). Clearly, these are substantial improvements compared
to the CR-EOMCC(2,3) computations, which confirm the usefulness of incorporating the leading triply excited determinants
identified by CIPSI in the \ccp\ and \eomccp\ calculations prior to the determination of the noniterative corrections
$\delta_{\mu}$($P$;$Q$). Tables\ \ref{table1} and \ref{table2} also show that the previously observed
\cite{cipsi-ccpq-2021,cbd_cipsi_ccpq} acceleration in the convergence of the ground-state \ccp\ energies
toward their CCSDT parents
offered by the \ccpq\ corrections applies to excited states and that the results of the CIPSI-driven \ccpq\ computations
systematically improve when the fractions of triples captured by CIPSI increase.
\subsection{\ce{CH}}
\label{sec3.3}
Our second example is the CH radical, where we examine energies of the ground and three lowest doublet excited states
at their respective experimentally derived equilibrium geometries used in Refs.\ \cite{crccl_ijqc2,hirata1} and our
previous semi-stochastic \ccpq\ work \cite{stochastic-ccpq-molphys-2020}, which are 1.1197868 \AA\ for the ${\rm X}\,^{2}\Pi$
ground state \cite{ch1}, 1.1031 \AA\ for the ${\rm A}\,^{2}\Delta$ state \cite{ch1}, 1.1640 \AA\ for the ${\rm B}\,^{2}\Sigma^{-}$
state \cite{ch2}, and 1.1143 \AA\ for the ${\rm C}\,^{2}\Sigma^{+}$ state \cite{ch3} (Table\ \ref{table3}).
In analogy to \ce{CH+}, the lists of triples defining the three-body components of the $T^{(P)}$,
$R_{\mu}^{(P)}$, and $L_{\mu}^{(P)}$ operators adopted in the CIPSI-driven \ccp, \eomccp, and \ccpq\
calculations for CH were extracted from the terminal $|\Psi^{\rm(CIPSI)}\rangle$ wave functions
representing the lowest-energy states of the relevant irreps of $C_{2v}$, {\it i.e.}, the
$^{2}{\rm B}_{2}(C_{2v})$ component of the ${\rm X}\,^{2}\Pi$ state, the lowest state of
$^{2}{\rm A}_{1}(C_{2v})$ symmetry for the ${\rm A}\,^{2}\Delta$ and ${\rm C}\,^{2}\Sigma^{+}$ states,
and the lowest $^{2}{\rm A}_{2}(C_{2v})$ state for the ${\rm B}\,^{2}\Sigma^{-}$ state.
The $\mathcal{V}_{\rm int}^{(0)}$ subspaces used to initiate the underlying CIPSI runs consisted of the
$^{2}{\rm B}_{2}(C_{2v})$-symmetric ROHF determinant for the ${\rm X}\,^{2}\Pi$ ground state,
the $^{2}{\rm A}_{1}(C_{2v})$-symmetric determinant of the $3\sigma\rightarrow1\pi$ type for the ${\rm A}\,^{2}\Delta$
and ${\rm C}\,^{2}\Sigma^{+}$ states, and the $3\sigma\rightarrow1\pi$/$3\sigma1\pi\rightarrow1\pi^{2}$
CSF of the $^{2}{\rm A}_{2}(C_{2v})$ symmetry for the ${\rm B}\,^{2}\Sigma^{-}$ state.

As shown in Refs.\ \cite{hirata1,crccl_ijqc2}, all three excited states of CH examined here,
especially B$\,^{2}\Sigma^{-}$ and C$\,^{2}\Sigma^{+}$ that are dominated by two-electron transitions,
represent a significant challenge. This is manifested by the large errors relative to EOMCCSDT obtained
for the ${\rm A}\,^{2}\Delta$, ${\rm B}\,^{2}\Sigma^{-}$, and ${\rm C}\,^{2}\Sigma^{+}$ states with EOMCCSD
(represented in Table\ \ref{table3} by \eomccp/$N_{\rm det(in)} = 1$), which are 13.474, 38.620, and
43.992 millihartree, respectively. The CR-EOMCC(2,3) triples corrections to EOMCCSD, equivalent to the
\ccpq/$N_{\rm det(in)} = 1$ calculations, reduce these errors, being very effective for the
${\rm C}\,^{2}\Sigma^{+}$ state, but the 7.727 millihartree error for the ${\rm A}\,^{2}\Delta$ state
and the 4.954 millihartree error for the ${\rm B}\,^{2}\Sigma^{-}$ state remain.
To address this situation, we turn to the CIPSI-driven \ccpq\ calculations with $N_{\rm det(in)} > 1$.
As shown in Table\ \ref{table3}, already the inexpensive \ccpq\ computations using $N_{\rm det(in)} = 1,000$,
which rely on the compact CIPSI wave functions spanned by at most 1,889 $S_{z} = 1/2$ determinants that capture
1.0--3.9\% of the 24,624--26,941 triples used by CCSDT/EOMCCSDT, reduce the 7.727 and 4.954 millihartree
errors relative to EOMCCSDT for the ${\rm A}\,^{2}\Delta$ and ${\rm B}\,^{2}\Sigma^{-}$ states to
1.779 and 0.239 millihartree, respectively. Upon further increasing $N_{\rm det(in)}$ to 5,000, which translates
into terminal CIPSI diagonalization spaces that for the four states of CH listed in Table\ \ref{table3} are 3--4
times smaller than the numbers of triples used by CCSDT/EOMCCSDT and into 13.0--18.7\% of all triples in the
resulting $P$ spaces, the CIPSI-driven \ccpq\ computations recover the parent CCSDT/EOMCCSDT energetics to within
0.051--0.367 millihartree, greatly improving the CR-CC(2,3)/CR-EOMCC(2,3) results for the ${\rm X}\,^{2}\Pi$,
${\rm A}\,^{2}\Delta$, and ${\rm B}\,^{2}\Sigma^{-}$ states, while retaining high accuracy for the
${\rm C}\,^{2}\Sigma^{+}$ state which CR-EOMCC(2,3) describes virtually perfectly. As in the case of
\ce{CH+}, the CIPSI-driven \ccpq\ calculations accelerate convergence of the underlying \ccp/\eomccp\ energetics
toward CCSDT/EOMCCSDT and become more accurate as the fractions of triples captured by CIPSI increase.
\subsection{Potential cuts of \ce{H2O}}
\label{sec3.4}
\begin{figure}[!ht]
\centering
\includegraphics[scale=0.38]{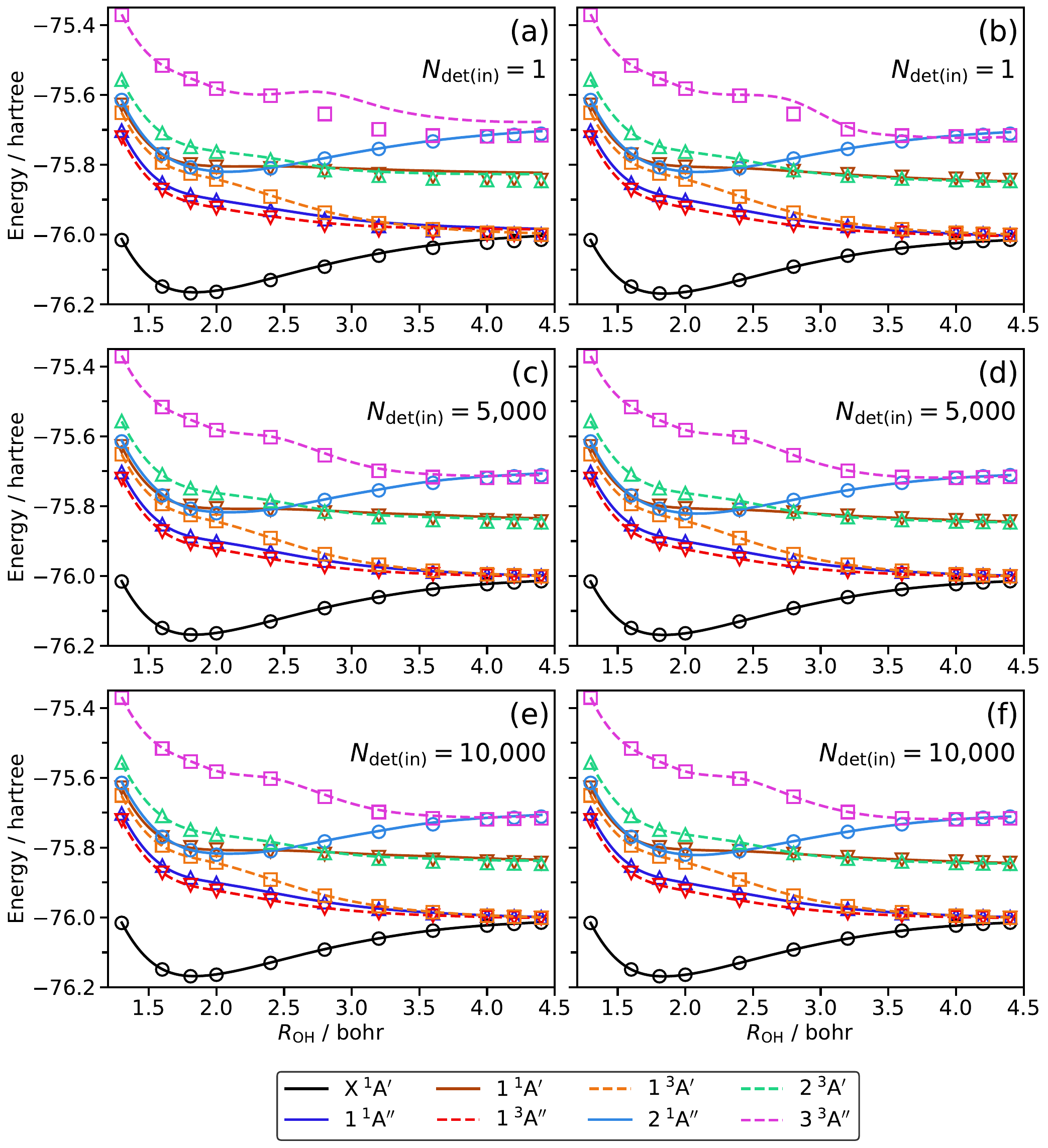}
\caption{
A comparison of the PES cuts of water along the O--H bond breaking coordinate $R_{\rm OH}$ for the
$\ce{H2O}\rightarrow{\rm H} + {\rm OH}$ dissociation channels correlating with the ${\rm X}\,^{2}\Pi$
ground state and the lowest-energy $^{2}\Sigma^{+}$ and $^{2}\Sigma^{-}$ states of the OH product resulting
from the CIPSI-driven \ccp/\eomccp\ 
[panels (a), (c), and (e)]
and \ccpq\
[panels (b), (d), and (f)]
calculations at selected values of the CIPSI wave function termination parameter $N_{\rm det(in)}$
with their full CCSDT/EOMCCSDT counterparts. The solid and dashed lines represent the splined
\ccp/\eomccp\ and \ccpq\ data, whereas the open circles, squares, triangles, and inverted triangles denote the
parent CCSDT/EOMCCSDT energetics. The \ccp/\eomccp\ and \ccpq\  results shown in panels (a) and (b),
obtained with $N_{\rm det(in)} = 1$, are equivalent to the CCSD/EOMCCSD and CR-CC(2,3)/CR-EOMCC(2,3) calculations,
respectively.
}
\label{fig1}
\end{figure}

Finally, we examine the ground- and excited-state PES cuts of water along the $\ce{H2O} \rightarrow \ce{H} + \ce{OH}$
dissociation path constructed using 11 values of the O--H bond separation $R_{\rm OH}$ spanning 1.3 to 4.4 bohr
and listed in Tables S1--S12 of the Supplementary Data document (Appendix A), with
the remaining O--H bond length and $\angle$H--O--H angle for each $R_{\rm OH}$ optimized using CCSD/cc-pVTZ
in Ref.\ \cite{nmapp3}. Following Ref. \cite{water} and our recent active-orbital-based and adaptive
\ccpq\ work \cite{kgjspp_h2o_2025}, we considered the four lowest \AP{}{}($C_{s}$)-symmetric singlet states
(the \AP{{\rm X}}{1} ground state and the excited \AP{n}{1} states with $n = 1\mbox{--}3$), the three lowest
\AP{}{}($C_{s}$)-symmetric triplet states (\AP{n}{3}, $n = 1\mbox{--}3$), the two lowest singlet states of the
\APP{}{}($C_{s}$) symmetry (\APP{n}{1}, $n = 1, 2$), and the three lowest \APP{}{}($C_{s}$)-symmetric triplet
states (\APP{n}{3}, $n = 1\mbox{--}3$). In analogy to \ce{CH+} and \ce{CH}, the lists of triples defining the
$T_{3}^{(P)}$, $R_{\mu,3}^{(P)}$, and $L_{\mu,3}^{(P)}$ components of the $T^{(P)}$, $R_{\mu}^{(P)}$, and
$L_{\mu}^{(P)}$ operators employed in the CIPSI-driven \ccp, \eomccp, and \ccpq\ calculations for \ce{H2O} were
extracted from the terminal $|\Psi^{\rm(CIPSI)}\rangle$ wave functions corresponding to the lowest-energy states
of the relevant symmetries, {\it i.e.}, we used the $S_{z}=0$ triples of the \AP{}{}($C_{s}$) symmetry for the
\AP{}{1} and \AP{}{3} states and the $S_{z}=0$ triples of the \APP{}{}($C_{s}$) symmetry for the \APP{}{1} and
\APP{}{3} states. The $\mathcal{V}_{\rm int}^{(0)}$ subspaces used to initiate the underlying CIPSI runs consisted
of the ${\rm RHF} = |(1a^{\prime})^{2}(2a^{\prime})^{2}(3a^{\prime})^{2}(4a^{\prime})^{2}(1a^{\prime\prime})^{2}|$
determinant for the \AP{}{} states and the $1a^{\prime\prime}\rightarrow5a^{\prime}$ CSFs for the \APP{}{} states.
The results of our computations are reported in Tables\ \ref{table4} and \ref{table5}, Fig.\ \ref{fig1}, and
Supplementary Data. Tables\ \ref{table4} and \ref{table5} summarize the mean unsigned error (MUE) and
nonparallelity error (NPE) values characterizing the PES cuts of water obtained in the CIPSI-driven \ccp/\eomccp\
and \ccpq\ calculations relative to their nearly exact \cite{kgjspp_h2o_2025} CCSDT/EOMCCSDT counterparts.
Figure\ \ref{fig1} shows the selected PES cuts associated with
the $\ce{H2O} \rightarrow \ce{H} + \ce{OH}$ dissociation channels that correlate with the ${\rm X}\,^{2}\Pi$ ground
state and the lowest $^{2}\Sigma^{+}$ and $^{2}\Sigma^{-}$ states of the OH product. The total electronic energies
for all the calculated states of water are provided in Supplementary Data.

In line with our earlier observations \cite{water,kgjspp_h2o_2025}, the CCSD/EOMCCSD and CR-CC(2,3)/CR-EOMCC(2,3)
approaches, represented in Tables\ \ref{table4} and \ref{table5}, Fig.\ \ref{fig1}, and
Supplementary Data by the $N_{\rm det(in)} = 1$ \ccp/\eomccp\ and \ccpq\ computations, respectively, behave well
near the equilibrium geometry on the ground-state PES. CR-CC(2,3) reduces the already small, 2.771--3.562
millihartree, errors in the CCSD energies relative to CCSDT for the \AP{{\rm X}}{1} ground state in the
$R_{\rm OH} = 1.3\mbox{--}2.0$ bohr region to 0.226--0.325 millihartree. For the excited states, the
largest deviation from EOMCCSDT obtained in this region with EOMCCSD is 3.392 millihartree (for the \AP{3}{1} state).
The largest error relative to EOMCCSDT produced in the same region by CR-EOMCC(2,3) is 1.660 millihartree
(for the \APP{3}{3} state). With the exception of the \APP{3}{3}, \AP{3}{1}, and \AP{3}{3} states, the
differences between the CR-EOMCC(2,3) and EOMCCSDT excited-state potentials in the $R_{\rm OH} = 1.3\mbox{--}2.0$
bohr region are $\sim$1 millihartree, confirming the utility of the CR-EOMCC(2,3) method in examining excited
states of water in the vicinity of its equilibrium geometry \cite{water}. The situation dramatically changes in
the $R_{\rm OH} > 2.0$ region, where nearly all electronic states of \ce{H2O} considered in this study acquire 
significant MR character, resulting in the large, 6.154--28.242 millihartree, MUEs and even larger, 8.453--64.664
millihartree, NPEs relative to CCSDT/EOMCCSDT obtained in the entire $R_{\rm OH} = 1.3\mbox{--}4.4$ bohr region for
the \AP{{\rm X}}{1}, \APP{1}{1}, \APP{1}{3}, \AP{1}{1}, \AP{2}{3}, \APP{2}{3}, \AP{2}{1}, \APP{3}{3}, \AP{3}{1}, and
\AP{3}{3} potentials with CCSD/EOMCCSD. The CR-CC(2,3)/CR-EOMCC(2,3)
triples corrections to CCSD/EOMCCSD reduce the MUE and NPE values relative to CCSDT/EOMCCSDT for all
12 potential cuts of water examined in this work to 0.543--6.054 and 0.442--41.943 millihartree, respectively, but
the substantial, 5.033 and 6.054 millihartree, MUEs and 13.044 and 41.943 millihartree
NPEs obtained for the \APP{2}{3} and \APP{3}{3} states remain. The failure of CR-EOMCC(2,3)
for the \APP{3}{3} state becomes particularly evident from the 36.791 millihartree error
relative to EOMCCSDT at $R_{\rm OH} = 2.8$ bohr, which results in a spurious bump in the \APP{3}{3}
potential and the massive NPE of 41.943 millihartree [see Fig.\ \ref{fig1}(b) and Table \ref{table5}].

To address the challenges encountered in the CR-EOMCC(2,3) calculations for some of the excited-state water
potentials examined in this work, we again turn to the CIPSI-driven \ccpq\ approach with $N_{\rm det(in)} > 1$.
As shown in Tables\ \ref{table4} and \ref{table5} and Tables S1--S12 in Supplementary Data, already
the \ccpq/$N_{\rm det(in)} = 1,000$ computations, which rely on the terminal $|\Psi^{\rm (CIPSI)}\rangle$ wave
functions obtained in the inexpensive CIPSI runs that contain only 1,021--1,994
$S_{z}=0$ determinants and 0.0--4.3\% of the 31,832 \AP{}{}($C_{s}$)-symmetric and 32,232
\APP{}{}($C_{s}$)-symmetric $S_{z}=0$ triples used by CCSDT/EOMCCSDT, provide substantial improvements,
reducing the 0.543--6.054 millihartree MUEs and 0.442--41.943 millihartree NPEs relative to CCSDT/EOMCCSDT
characterizing the CR-CC(2,3)/CR-EOMCC(2,3) potentials to 0.265--4.875 and 0.222--14.940 millihartree,
respectively. With a modest increase in computational
effort corresponding to $N_{\rm det(in)} = 5,000$, which results in 5,154--9,808 $S_{z}=0$ determinants of the
\AP{}{}($C_{s}$) and \APP{}{}($C_{s}$) symmetries in the final Hamiltonian diagonalization spaces adopted by CIPSI
and only 3.1--15.9\% of all triples in the respective $P$ spaces, the \ccpq\ calculations produce further improvements,
reproducing nearly all CCSDT/EOMCCSDT potentials considered in this study to within $\sim$1 millihartree. The accuracy
gains resulting from the \ccpq/$N_{\rm det(in)} = 5,000$ calculations are particularly impressive for the
aforementioned \APP{2}{3} and \APP{3}{3} states that are poorly described by CR-EOMCC(2,3) (and even worse
by the underlying EOMCCSD). In the case of the former state, the 5.033 millihartree MUE and 13.044 millihartree
NPE relative to EOMCCSDT obtained with CR-EOMCC(2,3) reduce in the \ccpq/$N_{\rm det(in)} = 5,000$ computations
to 0.695 and 0.559 millihartree, respectively. For the latter state, the \ccpq/$N_{\rm det(in)} = 5,000$
approach reduces the MUE and NPE values of 6.054 and 41.943 millihartree produced by CR-EOMCC(2,3) to 1.116 and
1.474 millihartee, respectively. In particular, the massive, 36.791 millihartree, error relative to EOMCCSDT
obtained for the \APP{3}{3} state at $R_{\rm OH} = 2.8$ bohr with CR-EOMCC(2,3) is reduced in the
\ccpq/$N_{\rm det(in)} = 5,000$ calculations to 1.680 millihartree, so that the unphysical bump on the
CR-EOMCC(2,3) [or \ccpq/$N_{\rm det(in)} = 1$] PES of the \APP{3}{3} state seen in the $R_{\rm OH} \approx 2.8$
bohr region in Fig.\ \ref{fig1}(b) disappears [see Fig.\ \ref{fig1}(d)]. The results for the water molecule
discussed here clearly demonstrate that the extension of our CIPSI-driven \ccpq\ methodology to excited states,
as presented in this work, is very promising, allowing us to accurately approximate the EOMCCSDT energetics, even
in challenging MR situations involving excited-state potentials along bond breaking coordinates, but there are
two high-lying states in Tables\ \ref{table4} and \ref{table5}, namely, \AP{3}{1} and \AP{3}{3}, for which the
convergence of the CIPSI-driven \ccpq\ energetics toward EOMCCSDT with $N_{\rm det(in)}$ is slow
(see, also, Tables S11 and S12 in Supplementary Data). This behavior arises because
in our current implementation of the CIPSI-driven \ccpq\ approach aimed at CCSDT/EOMCCSDT, the underlying
$P$ spaces are tailored to the lowest-energy states of the relevant symmetries (in the case of the
\AP{3}{1} and \AP{3}{3} states, the $S_{z} = 0$ determinants extracted from the CIPSI runs for the
\AP{{\rm X}}{1} ground state), which may be inadequate for describing higher-energy excited states, especially
when the fractions of triples captured by CIPSI are small. We will return to this issue in future work
by exploring state-specific $P$ spaces extracted from excited-state CIPSI calculations.
\section{Summary}
\label{sec4}
We extended the CIPSI-driven CC($P$;$Q$) methodology of Refs.\ \cite{cipsi-ccpq-2021,cbd_cipsi_ccpq}
to excited states. The resulting approach, aimed at converging CCSDT/EOMCCSDT energetics,
was tested on vertical excitations in \ce{CH+}, adiabatic excitations in CH,
and ground- and excited-state PES cuts of water along the O--H bond breaking coordinate.
We demonstrated that for nearly all electronic states examined, it is sufficient to use small fractions
of triply excited determinants in the underlying \ccp/\eomccp\ steps,
identified by CIPSI runs involving relatively inexpensive Hamiltonian diagonalizations,
to obtain CCSDT/EOMCCSDT-quality energetics, even when
MR correlations and $T_{3}$ and $R_{\mu,3}$ effects
are difficult to capture with the CR-CC(2,3)/CR-EOMCC(2,3) triples corrections to CCSD/EOMCCSD.
Among our future plans are the exploration of state-specific variants of the CIPSI-driven CC($P$;$Q$)
formalism
mentioned in Sec.\ \ref{sec2},
in which we will use CIPSI to design excitation spaces tailored to the electronic
states of interest, and the pursuit of further reductions in computational costs by replacing the
full CIPSI algorithm adopted in this work, which can explore the entire many-electron Hilbert space,
with its truncated analogs (such as CIPSI truncated at triples or triples and quadruples).
\section*{CRediT authorship contribution statement}
\textbf{Swati S. Priyadarsini}: Software, Data curation, Formal 
analysis, Investigation, Validation, Writing - original draft.
\textbf{Karthik Gururangan}: Methodology, Software, Data curation, Formal 
analysis, Investigation, Validation, Writing - reviewing and editing.
\textbf{Piotr Piecuch}: Conceptualization, Methodology, Formal 
analysis, Investigation, Funding acquisition, Project administration, 
Resources, Supervision, Validation, Writing - reviewing and editing.
\section*{Declaration of competing interest}
The authors declare that they have no known competing financial
interests or personal relationships that could have appeared to influence
the work reported in this paper.
\section*{Data availability}
The data that support the findings of this study are available
within the article and the Supplementary Data.
\section*{Acknowledgments}
This work has been supported by the Chemical Sciences,
Geosciences and Biosciences Division, 
Office of Basic Energy Sciences, 
Office of Science, 
U.S. Department of Energy 
(Grant No. DE-FG02-01ER15228 to P.P).
Support from the Jenison Fund at Michigan State University is gratefully
acknowledged too.
%
\section*{Appendix A. Supplementary data}
Supplementary material associated with this article can be found online at 

\bibliographystyle{elsarticle-num}
\biboptions{sort,compress}
\bibliography{refs}

\begin{thebibliography}{10}
\expandafter\ifx\csname url\endcsname\relax
  \def\url#1{\texttt{#1}}\fi
\expandafter\ifx\csname urlprefix\endcsname\relax\def\urlprefix{URL }\fi
\expandafter\ifx\csname href\endcsname\relax
  \def\href#1#2{#2} \def\path#1{#1}\fi

\bibitem{eomcc3}
J.~F. Stanton, R.~J. Bartlett, The equation of motion coupled-cluster method.
  {A} systematic biorthogonal approach to molecular excitation energies,
  transition probabilities, and excited state properties, J. Chem. Phys. 98
  (1993) 7029--7039.
\newblock \href {https://doi.org/10.1063/1.464746}
  {\path{doi:10.1063/1.464746}}.

\bibitem{cizek1}
J.~{\v C}{\'\i}{\v z}ek, On the correlation problem in atomic and molecular
  systems. {C}alculation of wavefunction components in {U}rsell-type expansion
  using quantum-field theoretical methods, J. Chem. Phys. 45 (1966) 4256--4266.
\newblock \href {https://doi.org/10.1063/1.1727484}
  {\path{doi:10.1063/1.1727484}}.

\bibitem{cizek4}
J.~Paldus, J.~{\v C}{\'\i}{\v z}ek, I.~Shavitt, Correlation problems in atomic
  and molecular systems. {IV}. {E}xtended coupled-pair many-electron theory and
  its application to the {BH}$_{3}$ molecule, Phys. Rev. A 5 (1972) 50--67.
\newblock \href {https://doi.org/10.1103/PhysRevA.5.50}
  {\path{doi:10.1103/PhysRevA.5.50}}.

\bibitem{monk}
H.~J. Monkhorst, Calculation of properties with the coupled-cluster method,
  Int. J. Quantum Chem., Symp. 11 (1977) 421--432.
\newblock \href {https://doi.org/10.1002/qua.560120850}
  {\path{doi:10.1002/qua.560120850}}.

\bibitem{sacci4}
H.~Nakatsuji, Cluster expansion of the wavefunction. {E}xcited states, Chem.
  Phys. Lett. 59 (1978) 362--364.
\newblock \href {https://doi.org/10.1016/0009-2614(78)89113-1}
  {\path{doi:10.1016/0009-2614(78)89113-1}}.

\bibitem{eomccsdt1}
K.~Kowalski, P.~Piecuch, The active-space equation-of-motion coupled-cluster
  methods for excited electronic states: {F}ull {EOMCCSDt}, J. Chem. Phys. 115
  (2001) 643--651.
\newblock \href {https://doi.org/10.1063/1.1378323}
  {\path{doi:10.1063/1.1378323}}.

\bibitem{eomccsdt2}
K.~Kowalski, P.~Piecuch, Excited-state potential energy curves of {CH}$^{+}$:
  {A} comparison of the {EOMCCSDt} and full {EOMCCSDT} results, Chem. Phys.
  Lett. 347 (2001) 237--246.
\newblock \href {https://doi.org/10.1016/S0009-2614(01)01010-7}
  {\path{doi:10.1016/S0009-2614(01)01010-7}}.

\bibitem{water}
J.~J. Lutz, P.~Piecuch, Performance of the completely renormalized
  equation-of-motion coupled-cluster method in calculations of excited-state
  potential cuts of water, Comput. Theor. Chem. 1040--1041 (2014) 20--34.
\newblock \href {https://doi.org/10.1016/j.comptc.2014.05.008}
  {\path{doi:10.1016/j.comptc.2014.05.008}}.

\bibitem{kgjspp_h2o_2025}
K.~Gururangan, J.~Shen, P.~Piecuch, Extension of the active-orbital-based and
  adaptive {CC($P$;$Q$)} approaches to excited electronic states: {A}pplication
  to potential cuts of water, Chem. Phys. Lett. 862 (2025) 141840.
\newblock \href {https://doi.org/https://doi.org/10.1016/j.cplett.2024.141840}
  {\path{doi:https://doi.org/10.1016/j.cplett.2024.141840}}.

\bibitem{creomcc-2015}
P.~Piecuch, J.~A. Hansen, A.~O. Ajala, Benchmarking the completely renormalised
  equation-of-motion coupled-cluster approaches for vertical excitation
  energies, Mol. Phys. 113 (2015) 3085--3127.
\newblock \href {https://doi.org/10.1080/00268976.2015.1076901}
  {\path{doi:10.1080/00268976.2015.1076901}}.

\bibitem{eomccsdt3}
S.~A. Kucharski, M.~W{\l}och, M.~Musia{\l}, R.~J. Bartlett, Coupled-cluster
  theory for excited electronic states: {T}he full equation-of-motion
  coupled-cluster single, double, and triple excitation method, J. Chem. Phys.
  115 (2001) 8263--8266.
\newblock \href {https://doi.org/10.1063/1.1416173}
  {\path{doi:10.1063/1.1416173}}.

\bibitem{eomap3}
J.~D. Watts, R.~J. Bartlett, Economical triple excitation equation-of-motion
  coupled-cluster methods for excitation energies, Chem. Phys. Lett. 233 (1995)
  81--87.
\newblock \href {https://doi.org/10.1016/0009-2614(94)01434-W}
  {\path{doi:10.1016/0009-2614(94)01434-W}}.

\bibitem{ccsdr3a}
O.~Christiansen, H.~Koch, P.~J{\o}rgensen, Perturbative triple excitation
  corrections to coupled cluster singles and doubles excitation energies, J.
  Chem. Phys. 105 (1996) 1451--1459.
\newblock \href {https://doi.org/10.1063/1.472007}
  {\path{doi:10.1063/1.472007}}.

\bibitem{eomap5}
J.~D. Watts, R.~J. Bartlett, Iterative and non-iterative triple excitation
  corrections in coupled-cluster methods for excited electronic states: {T}he
  {EOM-CCSDT}-3 and {EOM-CCSD($\rm{\tilde{T}}$)} methods, Chem. Phys. Lett. 258
  (1996) 581--588.
\newblock \href {https://doi.org/10.1016/0009-2614(96)00708-7}
  {\path{doi:10.1016/0009-2614(96)00708-7}}.

\bibitem{cc3_1}
O.~Christiansen, H.~Koch, P.~J{\o}rgensen, Response functions in the {CC3}
  iterative triple excitation model, J. Chem. Phys. 103 (1995) 7429--7441.
\newblock \href {https://doi.org/10.1063/1.470315}
  {\path{doi:10.1063/1.470315}}.

\bibitem{kkppeom}
K.~Kowalski, P.~Piecuch, New coupled-cluster methods with singles, doubles, and
  noniterative triples for high accuracy calculations of excited electronic
  states, J. Chem. Phys. 120 (2004) 1715--1738.
\newblock \href {https://doi.org/10.1063/1.1632474}
  {\path{doi:10.1063/1.1632474}}.

\bibitem{crccl_molphys}
M.~W{\l}och, M.~D. Lodriguito, P.~Piecuch, J.~R. Gour, Two new classes of
  non-iterative coupled-cluster methods derived from the method of moments of
  coupled-cluster equations, Mol. Phys. 104 (2006) 2149--2172, 104 (2006) 2991
  [Erratum].
\newblock \href {https://doi.org/10.1080/00268970600659586}
  {\path{doi:10.1080/00268970600659586}}.

\bibitem{crccl_ijqc2}
P.~Piecuch, J.~R. Gour, M.~W{\l}och, Left-eigenstate completely renormalized
  equation-of-motion coupled-cluster methods: {R}eview of key concepts,
  extension to excited states of open-shell systems, and comparison with
  electron-attached and ionized approaches, Int. J. Quantum Chem. 109 (2009)
  3268--3304.
\newblock \href {https://doi.org/10.1002/qua.22367}
  {\path{doi:10.1002/qua.22367}}.

\bibitem{7hq}
G.~Fradelos, J.~J. Lutz, T.~A. Weso{\l}owski, P.~Piecuch, M.~W{\l}och,
  Embedding vs supermolecular strategies in evaluating the
  hydrogen-bonding-induced shifts of excitation energies, J. Chem. Theory
  Comput. 7 (2011) 1647--1666.
\newblock \href {https://doi.org/10.1021/ct200101x}
  {\path{doi:10.1021/ct200101x}}.

\bibitem{jspp-chemphys2012}
J.~Shen, P.~Piecuch, Biorthogonal moment expansions in coupled-cluster theory:
  {R}eview of key concepts and merging the renormalized and active-space
  coupled-cluster methods, Chem. Phys. 401 (2012) 180--202.
\newblock \href {https://doi.org/10.1016/j.chemphys.2011.11.033}
  {\path{doi:10.1016/j.chemphys.2011.11.033}}.

\bibitem{eomccpt}
S.~Hirata, M.~Nooijen, I.~Grabowski, R.~J. Bartlett, Perturbative corrections
  to coupled-cluster and equation-of-motion coupled-cluster energies: {A}
  determinantal analysis, J. Chem. Phys. 114 (2001) 3919--3928, 115 (2001)
  3967--3968 [Erratum].
\newblock \href {https://doi.org/10.1063/1.1346578}
  {\path{doi:10.1063/1.1346578}}.

\bibitem{eomccpt2}
T.~Shiozaki, K.~Hirao, S.~Hirata, Second- and third-order triples and
  quadruples corrections to coupled-cluster singles and doubles in the ground
  and excited states, J. Chem. Phys. 126 (2007) 244106.
\newblock \href {https://doi.org/10.1063/1.2741262}
  {\path{doi:10.1063/1.2741262}}.

\bibitem{stochastic-ccpq-molphys-2020}
S.~H. Yuwono, A.~Chakraborty, J.~E. Deustua, J.~Shen, P.~Piecuch, Accelerating
  convergence of equation-of-motion coupled-cluster computations using the
  semi-stochastic {CC}(${P}$;${Q}$) formalism, Mol. Phys. 118 (2020) e1817592.
\newblock \href {https://doi.org/10.1080/00268976.2020.1817592}
  {\path{doi:10.1080/00268976.2020.1817592}}.

\bibitem{jspp-jcp2012}
J.~Shen, P.~Piecuch, Combining active-space coupled-cluster methods with moment
  energy corrections via the {CC}(${P}$;${Q}$) methodology, with benchmark
  calculations for biradical transition states, J. Chem. Phys. 136 (2012)
  144104.
\newblock \href {https://doi.org/10.1063/1.3700802}
  {\path{doi:10.1063/1.3700802}}.

\bibitem{jspp-jctc2012}
J.~Shen, P.~Piecuch, Merging active-space and renormalized coupled-cluster
  methods via the {CC}(${P}$;${Q}$) formalism, with benchmark calculations for
  singlet-triplet gaps in biradical systems, J. Chem. Theory Comput. 8 (2012)
  4968--4988.
\newblock \href {https://doi.org/10.1021/ct300762m}
  {\path{doi:10.1021/ct300762m}}.

\bibitem{nbjspp-molphys2017}
N.~P. Bauman, J.~Shen, P.~Piecuch, Combining active-space coupled-cluster
  approaches with moment energy corrections via the {CC}(${P}$;${Q}$)
  methodology: {C}onnected quadruple excitations, Mol. Phys. 115 (2017)
  2860--2891.
\newblock \href {https://doi.org/10.1080/00268976.2017.1350291}
  {\path{doi:10.1080/00268976.2017.1350291}}.

\bibitem{ccpq-be2-jpca-2018}
I.~Magoulas, N.~P. Bauman, J.~Shen, P.~Piecuch, Application of the
  {CC}(${P}$;${Q}$) hierarchy of coupled-cluster methods to the beryllium
  dimer, J. Phys. Chem. A 122 (2018) 1350--1368.
\newblock \href {https://doi.org/10.1021/acs.jpca.7b10892}
  {\path{doi:10.1021/acs.jpca.7b10892}}.

\bibitem{ccpq-mg2-mp-2019}
S.~H. Yuwono, I.~Magoulas, J.~Shen, P.~Piecuch, Application of the
  coupled-cluster {CC}(${P}$;${Q}$) approaches to the magnesium dimer, Mol.
  Phys. 117 (2019) 1486--1506.
\newblock \href {https://doi.org/10.1080/00268976.2018.1564847}
  {\path{doi:10.1080/00268976.2018.1564847}}.

\bibitem{stochastic-ccpq-prl-2017}
J.~E. Deustua, J.~Shen, P.~Piecuch, Converging high-level coupled-cluster
  energetics by {M}onte {C}arlo sampling and moment expansions, Phys. Rev.
  Lett. 119 (2017) 223003.
\newblock \href {https://doi.org/10.1103/PhysRevLett.119.223003}
  {\path{doi:10.1103/PhysRevLett.119.223003}}.

\bibitem{stochastic-ccpq-jcp-2021}
J.~E. Deustua, J.~Shen, P.~Piecuch, High-level coupled-cluster energetics by
  {M}onte {C}arlo sampling and moment expansions: {F}urther details and
  comparisons, J. Chem. Phys. 154 (2021) 124103.
\newblock \href {https://doi.org/10.1063/5.0045468}
  {\path{doi:10.1063/5.0045468}}.

\bibitem{cipsi-ccpq-2021}
K.~Gururangan, J.~E. Deustua, J.~Shen, P.~Piecuch, High-level coupled-cluster
  energetics by merging moment expansions with selected configuration
  interaction, J. Chem. Phys. 155 (2021) 174114.
\newblock \href {https://doi.org/10.1063/5.0064400}
  {\path{doi:10.1063/5.0064400}}.

\bibitem{arnab-stgap-2022}
A.~Chakraborty, S.~H. Yuwono, J.~E. Deustua, J.~Shen, P.~Piecuch, Benchmarking
  the semi-stochastic {CC}(${P}$;${Q}$) approach for singlet--triplet gaps in
  biradicals, J. Chem. Phys. 157 (2022) 134101.
\newblock \href {https://doi.org/10.1063/5.0100165}
  {\path{doi:10.1063/5.0100165}}.

\bibitem{adaptiveccpq2023}
K.~Gururangan, P.~Piecuch, Converging high-level coupled-cluster energetics via
  adaptive selection of excitation manifolds driven by moment expansions, J.
  Chem. Phys. 159 (2023) 084108.
\newblock \href {https://doi.org/10.1063/5.0162873}
  {\path{doi:10.1063/5.0162873}}.

\bibitem{cbd_cipsi_ccpq}
S.~S. Priyadarsini, K.~Gururangan, J.~Shen, P.~Piecuch, The singlet--triplet
  gap of cyclobutadiene: {T}he {CIPSI}-driven {CC($P$;$Q$)} study, J. Phys.
  Chem. A 129 (2025) 11749--11780.
\newblock \href {https://doi.org/https://doi.org/10.1021/acs.jpca.5c07572}
  {\path{doi:https://doi.org/10.1021/acs.jpca.5c07572}}.

\bibitem{crccl_jcp}
P.~Piecuch, M.~W{\l}och, Renormalized coupled-cluster methods exploiting left
  eigenstates of the similarity-transformed {H}amiltonian, J. Chem. Phys. 123
  (2005) 224105.
\newblock \href {https://doi.org/10.1063/1.2137318}
  {\path{doi:10.1063/1.2137318}}.

\bibitem{ccfullt}
J.~Noga, R.~J. Bartlett, The full {CCSDT} model for molecular electronic
  structure, J. Chem. Phys. 86 (1987) 7041--7050, 89 (1988) 3401 [Erratum].
\newblock \href {https://doi.org/10.1063/1.452353}
  {\path{doi:10.1063/1.452353}}.

\bibitem{ccfullt2}
G.~E. Scuseria, H.~F. Schaefer, III, A new implementation of the full {CCSDT}
  model for molecular electronic structure, Chem. Phys. Lett. 152 (1988)
  382--386.
\newblock \href {https://doi.org/10.1016/0009-2614(88)80110-6}
  {\path{doi:10.1016/0009-2614(88)80110-6}}.

\bibitem{sci_3}
B.~Huron, J.~P. Malrieu, P.~Rancurel, Iterative perturbation calculations of
  ground and excited state energies from multiconfigurational zeroth-order
  wavefunctions, J. Chem. Phys. 58 (1973) 5745--5759.
\newblock \href {https://doi.org/https://doi.org/10.1063/1.1679199}
  {\path{doi:https://doi.org/10.1063/1.1679199}}.

\bibitem{cipsi_1}
Y.~Garniron, A.~Scemama, P.-F. Loos, M.~Caffarel, Hybrid
  stochastic-deterministic calculation of the second-order perturbative
  contribution of multireference perturbation theory, J. Chem. Phys. 147 (2017)
  034101.
\newblock \href {https://doi.org/https://doi.org/10.1063/1.4992127}
  {\path{doi:https://doi.org/10.1063/1.4992127}}.

\bibitem{cipsi_2}
Y.~Garniron, T.~Applencourt, K.~Gasperich, A.~Benali, A.~Fert\'{e}, J.~Paquier,
  B.~Pradines, R.~Assaraf, P.~Reinhardt, J.~Toulouse, P.~Barbaresco, N.~Renon,
  G.~David, J.-P. Malrieu, M.~V\'{e}ril, M.~Caffarel, P.-F. Loos, E.~Giner,
  A.~Scemama, Quantum {P}ackage 2.0: {A}n open-source determinant-driven suite
  of programs, J. Chem. Theory Comput. 15 (2019) 3591--3609.
\newblock \href {https://doi.org/https://doi.org/10.1021/acs.jctc.9b00176}
  {\path{doi:https://doi.org/10.1021/acs.jctc.9b00176}}.

\bibitem{chplusbasis}
J.~Olsen, A.~M. {S{\'a}nchez de Mer{\'a}s}, H.~J.~A. Jensen, P.~J{\o}rgensen,
  Excitation energies, transition moments and dynamic polarizabilities for
  \ce{CH+}. {A} comparison of multiconfigurational linear response and full
  configuration interaction calculations, Chem. Phys. Lett. 154 (1989)
  380--386.
\newblock \href {https://doi.org/10.1016/0009-2614(89)85373-4}
  {\path{doi:10.1016/0009-2614(89)85373-4}}.

\bibitem{ccpvnz}
T.~H. Dunning, Jr., Gaussian basis sets for use in correlated molecular
  calculations. {I.} {T}he atoms boron through neon and hydrogen, J. Chem.
  Phys. 90 (1989) 1007--1023.
\newblock \href {https://doi.org/10.1063/1.456153}
  {\path{doi:10.1063/1.456153}}.

\bibitem{augccpvnz}
R.~A. Kendall, T.~H. Dunning, Jr., R.~J. Harrison, Electron affinities of the
  first-row atoms revisited. {S}ystematic basis sets and wave functions, J.
  Chem. Phys. 96 (1992) 6796--6806.
\newblock \href {https://doi.org/10.1063/1.462569}
  {\path{doi:10.1063/1.462569}}.

\bibitem{hirata1}
S.~Hirata, Higher-order equation-of-motion coupled-cluster methods, J. Chem.
  Phys. 121 (2004) 51--59.
\newblock \href {https://doi.org/10.1063/1.1753556}
  {\path{doi:10.1063/1.1753556}}.

\bibitem{nmapp3}
X.~Li, J.~Paldus, Performance of multireference and equation-of-motion
  coupled-cluster methods for potential energy surfaces of low-lying excited
  states: {S}ymmetric and asymmetric dissociation of water, J. Chem. Phys. 133
  (2010) 024102.
\newblock \href {https://doi.org/10.1063/1.3451074}
  {\path{doi:10.1063/1.3451074}}.

\bibitem{CCpy-GitHub}
K. Gururangan, J. E. Deustua, and P. Piecuch, ``CCpy: A Coupled-Cluster Package
  Written in Python,'' see https://github.com/piecuch-group/ccpy [software].

\bibitem{gamess2020}
G.~M.~J. Barca, C.~Bertoni, L.~Carrington, D.~Datta, N.~De~Silva, J.~E.
  Deustua, D.~G. Fedorov, J.~R. Gour, A.~O. Gunina, E.~Guidez, T.~Harville,
  S.~Irle, J.~Ivanic, K.~Kowalski, S.~S. Leang, H.~Li, W.~Li, J.~J. Lutz,
  I.~Magoulas, J.~Mato, V.~Mironov, H.~Nakata, B.~Q. Pham, P.~Piecuch,
  D.~Poole, S.~R. Pruitt, A.~P. Rendell, L.~B. Roskop, K.~Ruedenberg,
  T.~Sattasathuchana, M.~W. Schmidt, J.~Shen, L.~Slipchenko, M.~Sosonkina,
  V.~Sundriyal, A.~Tiwari, J.~L.~G. Vallejo, B.~Westheimer, M.~W{\l}och, P.~Xu,
  F.~Zahariev, M.~S. Gordon, Recent developments in the general atomic and
  molecular electronic structure system, J. Chem. Phys. 152 (2020) 154102.
\newblock \href {https://doi.org/10.1063/5.0005188}
  {\path{doi:10.1063/5.0005188}}.

\bibitem{ch1}
M.~Zachwieja, New investigations of the {A $^2\Delta$ -- X $^2\Pi$} band system
  in the {CH} radical and a new reduction of the vibration-rotation spectrum of
  {CH} from the {ATMOS} spectra, J. Mol. Spectrosc. 170 (1995) 285--309.
\newblock \href {https://doi.org/10.1006/jmsp.1995.1072}
  {\path{doi:10.1006/jmsp.1995.1072}}.

\bibitem{ch2}
R.~K{\c e}pa, A.~Para, M.~Rytel, M.~Zachwieja, New spectroscopic analysis of
  the {B $^2\Sigma^-$ -- X $^2\Pi$} band system of the {CH} molecule, J. Mol.
  Spectrosc. 178 (1996) 189--193.
\newblock \href {https://doi.org/10.1006/jmsp.1996.0173}
  {\path{doi:10.1006/jmsp.1996.0173}}.

\bibitem{ch3}
K.~P. Huber, G.~Herzberg, Molecular Spectra and Molecular Structure: {IV.}
  Constants of Diatomic Molecules, Van Nostrand Reinhold, New York, 1979.
\newblock \href {https://doi.org/10.1007/978-1-4757-0961-2}
  {\path{doi:10.1007/978-1-4757-0961-2}}.

\end{thebibliography}

\newpage
\clearpage
\pagebreak
\begin{sidewaystable}
\centering
\footnotesize
\begin{threeparttable}
\caption{Convergence of the CIPSI-based \ccp/\eomccp\ and \ccpq\ energies toward CCSDT/EOMCCSDT
for the \ce{CH+} ion, as described by the [5s3p1d/3s1p] basis set of Ref.\ \cite{chplusbasis}, at the
C--H internuclear distance $R = R_{\rm e} = 2.13713$ bohr. The $P$ spaces used in the \ccp\ and
\eomccp\ calculations were defined as all singly and doubly excited determinants and subsets of triply
excited determinants extracted from the terminal $|\Psi^{\rm(CIPSI)}\rangle$ wave functions obtained in
the CIPSI runs for the lowest-energy states of the relevant symmetries, as described in the main text
(see, also, footnote `a'). The $Q$ spaces used to construct the \ccpq\ corrections $\delta_{\mu}(P;\!Q)$
consisted of the remaining triples not captured by CIPSI.}
\label{table1}
\renewcommand{\arraystretch}{1.5}
\setlength{\tabcolsep}{1.5pt}
\begin{tabular}{
c @{\extracolsep{0.12in}}
c @{\extracolsep{0.10in}} c @{\extracolsep{0.10in}} c @{\extracolsep{0.10in}} c @{\extracolsep{0.12in}}
c @{\extracolsep{0.10in}} c @{\extracolsep{0.12in}}
c @{\extracolsep{0.10in}} c @{\extracolsep{0.12in}}
c @{\extracolsep{0.10in}} c @{\extracolsep{0.12in}}
c @{\extracolsep{0.10in}} c @{\extracolsep{0.10in}} c @{\extracolsep{0.10in}} c @{\extracolsep{0.12in}}
c @{\extracolsep{0.10in}} c @{\extracolsep{0.12in}}
c @{\extracolsep{0.10in}} c @{\extracolsep{0.10in}} c @{\extracolsep{0.10in}} c @{\extracolsep{0.12in}}
c @{\extracolsep{0.10in}} c}
\hline\hline
\multirow{2}{*}{$N_{\rm det(in)}$}
& \multicolumn{4}{c}{$1\,^1\Sigma^{+}$}
& \multicolumn{2}{c}{$2\,^1\Sigma^{+}$}
& \multicolumn{2}{c}{$3\,^1\Sigma^{+}$}
& \multicolumn{2}{c}{$4\,^1\Sigma^{+}$}
& \multicolumn{4}{c}{$1\,^1\Pi$}
& \multicolumn{2}{c}{$2\,^1\Pi$}
& \multicolumn{4}{c}{$1\,^1\Delta$}
& \multicolumn{2}{c}{$2\,^1\Delta$} \\
\cline{2-5} \cline{6-7} \cline{8-9} \cline{10-11}
\cline{12-15} \cline{16-17} \cline{18-21} \cline{22-23}
{ }
& $N_{\rm det(out)}$\tnote{a} & \%T\tnote{b} & $P$\tnote{c} & ($P$;$Q$)\tnote{d}
& $P$\tnote{c} & ($P$;$Q$)\tnote{d}
& $P$\tnote{c} & ($P$;$Q$)\tnote{d}
& $P$\tnote{c} & ($P$;$Q$)\tnote{d}
& $N_{\rm det(out)}$\tnote{a} & \%T\tnote{b} & $P$\tnote{c} & ($P$;$Q$)\tnote{d}
& $P$\tnote{c} & ($P$;$Q$)\tnote{d}
& $N_{\rm det(out)}$\tnote{a} & \%T\tnote{b} & $P$\tnote{c} & ($P$;$Q$)\tnote{d}
& $P$\tnote{c} & ($P$;$Q$)\tnote{d} \\
\hline
1\tnote{e} & 1       & 0.0  & 1.845 & 0.063 & 19.694 & 1.373 & 3.856    & 0.787 & 5.536 & 0.954 & 2      & 0.0  & 3.080 & 0.792 & 11.656 & 2.805 & 2      & 0.0  & 34.304 & $-0.499$ & 34.685 & 0.350 \\
1,000      & 1,347   & 0.7  & 0.932 & 0.021 & 12.386 & 0.808 & 3.127    & 0.599 & 2.902 & 0.112 & 1,286  & 2.9  & 0.769 & 0.189 & 3.825  & 0.545 & 1,178  & 2.6  & 2.084  & 0.181    & 10.072 & 0.467 \\ 
5,000      & 5,392   & 5.7  & 0.171 & 0.005 & 4.117  & 0.207 & 0.325    & 0.066 & 0.479 & 0.061 & 6,610  & 13.9 & 0.138 & 0.044 & 2.041  & 0.154 & 9,446  & 16.8 & 0.087  & 0.023    & 1.082  & 0.066 \\
10,000     & 10,798  & 11.6 & 0.062 & 0.001 & 0.889  & 0.062 & $-0.039$ & 0.043 & 0.256 & 0.032 & 13,226 & 23.3 & 0.043 & 0.020 & 1.190  & 0.042 & 19,156 & 26.9 & 0.026  & 0.009    & 0.317  & 0.018 \\
20,000     & 21,602  & 21.4 & 0.019 & 0.000 & 0.201  & 0.015 & $-0.115$ & 0.019 & 0.161 & 0.013 & 23,756 & 34.0 & 0.018 & 0.009 & 0.181  & 0.012 & 32,778 & 36.7 & 0.010  & 0.004    & 0.179  & 0.006 \\
\hline\hline
\end{tabular}
\begin{tablenotes}
\footnotesize
\item[a]{
For each value of $N_{\rm det(in)}$, $N_{\rm det(out)}$ is the total number of determinants in the corresponding
terminal $|\Psi^{\rm(CIPSI)}\rangle$ wave function obtained in a CIPSI run for the lowest state of a given symmetry
[the $1 \,^{1}\Sigma^{+} = 1 \,^{1}{\rm A}_{1}(C_{2v})$ ground state for the $^{1}\Sigma^{+}$ states, the
$^{1}{\rm B}_{1}(C_{2v})$ component of the $1 \,^{1}\Pi$ state for the $^{1}\Pi$ states, and the
$^{1}{\rm A}_{2}(C_{2v})$ component of the $1 \,^{1}\Delta$ state for the $^{1}\Delta$ states].}
\item[b]{
The \%T values are the percentages of triply excited determinants contained in the terminal CIPSI wave
functions $|\Psi^{\rm(CIPSI)}\rangle$ described in footnote `a'.}
\item[c]{
Errors in the \ccp\ (the 1$\,^{1}\Sigma^{+}$ ground state) and \eomccp\ (excited states) energies,
in millihartree, relative to the CCSDT and EOMCCSDT data.
The CCSDT energy for the $1\,^{1}\Sigma^{+}$ state is
$-38.019516$ hartree. The EOMCCSDT energies for the $2\,^{1}\Sigma^{+}$, $3\,^{1}\Sigma^{+}$,
$4\,^{1}\Sigma^{+}$, $1\,^{1}\Pi$, $2\,^{1}\Pi$, $1\,^{1}\Delta$, and $2\,^{1}\Delta$ states
are $-37.702621$, $-37.522457$, $-37.386872$, $-37.900921$, $-37.498143$, $-37.762113$, and
$-37.402308$ hartree, respectively.}
\item[d]{
Errors, in millihartree, in the \ccpq\ energies relative to the corresponding CCSDT and EOMCCSDT
data provided in footnote `c'.}
\item[e]{
The \ccp\ and \eomccp\ energies at $N_{\rm det(in)} = 1$ are identical to the energies obtained in
the CCSD and EOMCCSD calculations. The corresponding \ccpq\ energies are equivalent to the CR-CC(2,3)
(the ground state) and CR-EOMCC(2,3) (excited states) results.}
\end{tablenotes}
\end{threeparttable}
\end{sidewaystable}

\newpage
\clearpage
\pagebreak
\begin{sidewaystable}
\centering
\footnotesize
\begin{threeparttable}
\caption{Same as Table \ref{table1} for the stretched C--H internuclear distance
$R = 2R_{\rm e} = 4.27426$ bohr.\tnote{a}}
\label{table2}
\renewcommand{\arraystretch}{1.5}
\setlength{\tabcolsep}{1.5pt}
\begin{tabular}{
c @{\extracolsep{0.12in}}
c @{\extracolsep{0.10in}} c @{\extracolsep{0.10in}} c @{\extracolsep{0.10in}} c @{\extracolsep{0.12in}}
c @{\extracolsep{0.10in}} c @{\extracolsep{0.12in}}
c @{\extracolsep{0.10in}} c @{\extracolsep{0.12in}}
c @{\extracolsep{0.10in}} c @{\extracolsep{0.12in}}
c @{\extracolsep{0.10in}} c @{\extracolsep{0.10in}} c @{\extracolsep{0.10in}} c @{\extracolsep{0.12in}}
c @{\extracolsep{0.10in}} c @{\extracolsep{0.12in}}
c @{\extracolsep{0.10in}} c @{\extracolsep{0.10in}} c @{\extracolsep{0.10in}} c @{\extracolsep{0.12in}}
c @{\extracolsep{0.10in}} c}
\hline\hline
\multirow{2}{*}{$N_{\rm det(in)}$}
& \multicolumn{4}{c}{$1\,^1\Sigma^{+}$}
& \multicolumn{2}{c}{$2\,^1\Sigma^{+}$}
& \multicolumn{2}{c}{$3\,^1\Sigma^{+}$}
& \multicolumn{2}{c}{$4\,^1\Sigma^{+}$}
& \multicolumn{4}{c}{$1\,^1\Pi$}
& \multicolumn{2}{c}{$2\,^1\Pi$}
& \multicolumn{4}{c}{$1\,^1\Delta$}
& \multicolumn{2}{c}{$2\,^1\Delta$} \\
\cline{2-5} \cline{6-7} \cline{8-9} \cline{10-11}
\cline{12-15} \cline{16-17} \cline{18-21} \cline{22-23}
{ }
& $N_{\rm det(out)}$ & \%T & $P$ & ($P$;$Q$)
& $P$ & ($P$;$Q$)
& $P$ & ($P$;$Q$)
& $P$ & ($P$;$Q$)
& $N_{\rm det(out)}$ & \%T & $P$ & ($P$;$Q$)
& $P$ & ($P$;$Q$)
& $N_{\rm det(out)}$ & \%T & $P$ & ($P$;$Q$)
& $P$ & ($P$;$Q$) \\
\hline
1      & 1      & 0.0  & 5.002 & 0.012 & 17.140 & 1.646 & 19.929 & $-2.870$ & 32.639 & 12.657 & 2      & 0.0  & 13.552 & 2.303 & 21.200 & $-1.428$ & 2       & 0.0  & 44.495 & $-4.525$ & 144.414 & $-63.405$ \\
1,000  & 1,677  & 1.8  & 0.434 & 0.005 & 7.457  & 1.096 & 10.848 & $-1.282$ & 9.609  & 0.999  & 1,792  & 3.2  & 0.452  & 0.046 & 1.758  & 0.095    & 1,764   & 3.7  & 0.430  & 0.046    & 8.160   & 0.855     \\
5,000  & 6,719  & 7.7  & 0.060 & 0.001 & 0.852  & 0.034 & 0.889  & 0.011    & 1.860  & 0.109  & 7,180  & 11.0 & 0.052  & 0.006 & 0.159  & 0.012    & 6,962   & 10.8 & 0.047  & 0.008    & 0.408   & 0.011     \\
10,000 & 13,163 & 15.3 & 0.016 & 0.000 & 0.404  & 0.010 & 0.446  & 0.008    & 0.760  & 0.055  & 14,362 & 19.2 & 0.013  & 0.002 & 0.043  & 0.003    & 13,312  & 17.4 & 0.015  & 0.004    & 0.144   & 0.002     \\
20,000 & 24,508 & 24.7 & 0.003 & 0.000 & 0.098  & 0.005 & 0.096  & 0.005    & 0.148  & 0.007  & 27,560 & 29.6 & 0.003  & 0.001 & 0.015  & 0.003    & 28,274  & 28.1 & 0.004  & 0.002    & 0.037   & 0.000     \\
\hline\hline
\end{tabular}
\begin{tablenotes}
\footnotesize
\item[a]{
The CCSDT energy for the $1\,^{1}\Sigma^{+}$ state is $-37.900394$ hartree. The EOMCCSDT
energies for the $2\,^{1}\Sigma^{+}$, $3\,^{1}\Sigma^{+}$, $4\,^{1}\Sigma^{+}$, $1\,^{1}\Pi$,
$2\,^{1}\Pi$, $1^{1}\Delta$, and $2^{1}\Delta$ states are $-37.704834$, $-37.650242$,
$-37.495275$, $-37.879532$, $-37.702345$, $-37.714180$, and $-37.494031$ hartree, respectively.}
%
\end{tablenotes}
\end{threeparttable}
\end{sidewaystable}

\newpage
\clearpage
\pagebreak
\begin{sidewaystable}
\centering
\footnotesize
\begin{threeparttable}
\caption{Convergence of the CIPSI-based \ccp/\eomccp\ and \ccpq\ energies toward CCSDT/EOMCCSDT
for the CH radical, as described by the aug-cc-pVDZ basis set \cite{ccpvnz,augccpvnz}. The $P$
spaces used in the \ccp\ and \eomccp\ calculations were defined as all singly and doubly excited
determinants and subsets of triply excited determinants extracted from the terminal $|\Psi^{\rm(CIPSI)}\rangle$
wave functions obtained in the CIPSI runs for the lowest-energy states of the relevant symmetries,
as described in the main text (see, also, footnote `a'). The $Q$ spaces used to construct the \ccpq\
corrections $\delta_{\mu}(P;\!Q)$ consisted of the remaining triples not captured by CIPSI.}
\label{table3}
\renewcommand{\arraystretch}{1.5}
\setlength{\tabcolsep}{1.5pt}
\begin{tabular}{
c @{\extracolsep{0.20in}}
c @{\extracolsep{0.10in}} c @{\extracolsep{0.10in}}
c @{\extracolsep{0.10in}} c @{\extracolsep{0.25in}}
c @{\extracolsep{0.10in}} c @{\extracolsep{0.10in}}
c @{\extracolsep{0.10in}} c @{\extracolsep{0.25in}}
c @{\extracolsep{0.10in}} c @{\extracolsep{0.10in}}
c @{\extracolsep{0.10in}} c @{\extracolsep{0.25in}}
c @{\extracolsep{0.10in}} c @{\extracolsep{0.10in}}
c @{\extracolsep{0.10in}} c}
\hline\hline
{ }
& \multicolumn{4}{c}{${\rm X}\,^2\Pi$}
& \multicolumn{4}{c}{${\rm A}\,^2\Delta$}
& \multicolumn{4}{c}{${\rm B}\,^2\Sigma^-$}
& \multicolumn{4}{c}{${\rm C}\,^2\Sigma^+$} \\
\cline{2-5}
\cline{6-9}
\cline{10-13}
\cline{14-17}
$N_{\rm det(in)}$
& $N_{\rm det(out)}$\tnote{a} & \%T\tnote{b} & $P$\tnote{c} & ($P$;$Q$)\tnote{d}
& $N_{\rm det(out)}$\tnote{a} & \%T\tnote{b} & $P$\tnote{c} & ($P$;$Q$)\tnote{d}
& $N_{\rm det(out)}$\tnote{a} & \%T\tnote{b} & $P$\tnote{c} & ($P$;$Q$)\tnote{d}
& $N_{\rm det(out)}$\tnote{a} & \%T\tnote{b} & $P$\tnote{c} & ($P$;$Q$)\tnote{d} \\
\hline
1\tnote{e} & 1      & 0.0  & 2.987 & 0.231 & 1      & 0.0  & 13.474 & 7.727 & 3      & 0.0  & 38.620 & $-4.954$ & 1      & 0.0  & 43.992 & 0.087 \\
1,000      & 1,154  & 1.0  & 2.909 & 0.227 & 1,677  & 3.3  & 6.482  & 1.779 & 1,790  & 3.9  & 5.914  & 0.239    & 1,889  & 3.9  & 10.044 & 0.922 \\
5,000      & 9,247  & 16.4 & 0.917 & 0.051 & 6,586  & 13.0 & 1.410  & 0.266 & 7,184  & 18.7 & 2.167  & 0.367    & 7,567  & 14.7 & 1.747  & 0.309 \\
10,000     & 18,501 & 34.0 & 0.418 & 0.019 & 13,425 & 24.4 & 0.585  & 0.115 & 14,373 & 34.6 & 1.172  & 0.204    & 11,014 & 20.3 & 1.310  & 0.209 \\
20,000     & 37,016 & 62.5 & 0.096 & 0.003 & 26,851 & 43.4 & 0.190  & 0.034 & 28,755 & 54.9 & 0.569  & 0.080    & 22,870 & 38.5 & 0.690  & 0.091 \\
\hline\hline
\end{tabular}
\begin{tablenotes}
\footnotesize
\item[a]{
For each value of $N_{\rm det(in)}$, $N_{\rm det(out)}$ is the total number of determinants in the corresponding
terminal $|\Psi^{\rm(CIPSI)}\rangle$ wave function obtained in a CIPSI run for the lowest state of a given symmetry
[the $^{2}{\rm B}_{2}(C_{2v})$ component of the ${\rm X} \,^{2}\Pi$ ground state, the lowest $^{2}{\rm A}_{1}(C_{2v})$
state for the ${\rm A} \,^{2}\Delta$ and ${\rm C} \,^{2}\Sigma^{+}$ states, and the lowest $^{2}{\rm A}_{2}(C_{2v})$
state for the ${\rm B} \,^{2}\Sigma^{-}$ state].}
\item[b]{
The \%T values are the percentages of triply excited determinants contained in the terminal CIPSI wave
functions $|\Psi^{\rm(CIPSI)}\rangle$ described in footnote `a'.}
\item[c]{
Errors in the \ccp\ (the X$\,^{2}\Pi$ ground state) and \eomccp\ (excited states) energies,
in millihartree, relative to the CCSDT and EOMCCSDT data,
calculated at the experimentally derived equilibrium C--H bond lengths
used in Refs.\ \cite{hirata1,crccl_ijqc2}, which are 1.1197868 \AA\ for the
${\rm X} \,^{2}\Pi$ state \cite{ch1}, 1.1031 \AA\ for the ${\rm A} \,^{2}\Delta$ state \cite{ch1},
1.1640 \AA\ for the ${\rm B} \,^{2}\Sigma^{-}$ state \cite{ch2}, and 1.1143 \AA\ for the
${\rm C} \,^{2}\Sigma^{+}$ state \cite{ch3}. The CCSDT and EOMCCSDT energies for the
${\rm X} \,^{2}\Pi$, ${\rm A} \,^{2}\Delta$, ${\rm B} \,^{2}\Sigma^{-}$, and ${\rm C} \,^{2}\Sigma^{+}$
states at these geometries are $-38.387749$, $-38.276770$, $-38.267544$, and $-38.238205$ hartree, respectively.}
\item[d]{
Errors, in millihartree, in the \ccpq\ energies relative to the corresponding CCSDT and EOMCCSDT
data, obtained at the experimentally derived equilibrium C--H bond lengths used in Refs.\
\cite{hirata1,crccl_ijqc2} and listed in footnote `c'.}
\item[e]{
The \ccp\ and \eomccp\ energies at $N_{\rm det(in)} = 1$ are identical to the energies obtained in
the CCSD and EOMCCSD calculations. The corresponding \ccpq\ energies are equivalent to the CR-CC(2,3)
(the ground state) and CR-EOMCC(2,3) (excited states) results.}
\end{tablenotes}
\end{threeparttable}
\end{sidewaystable}

\newpage
\clearpage
\pagebreak
\begin{table*}[ht!]
\centering
\footnotesize
\begin{threeparttable}
\caption{The MUE values, in millihartree, relative to CCSDT/EOMCCSDT characterizing the
ground- and 11 excited-state potential cuts of the water molecule, as described by the TZ
basis set of Ref.\ \cite{nmapp3}, along the O--H bond-breaking coordinate corresponding
to the $\ce{H2O} \rightarrow {\rm H} + {\rm OH}$ dissociation obtained with the CIPSI-based
\ccp/\eomccp\ and \ccpq\ approaches examined in the present work. The $P$ spaces used in
the \ccp\ and \eomccp\ calculations were defined as all singly and doubly excited
determinants and subsets of triply excited determinants extracted from the terminal $|\Psi^{\rm(CIPSI)}\rangle$
wave functions obtained in the CIPSI runs for the lowest-energy states of the relevant symmetries,
as described in the main text (see, also, footnote `b'). The $Q$ spaces used to construct the \ccpq\
corrections $\delta_{\mu}(P;\!Q)$ consisted of the remaining triples not captured by CIPSI.}
\label{table4}
\renewcommand{\arraystretch}{1.5}
\setlength{\tabcolsep}{1.2pt}
\begin{tabular}{
C{1.2cm} @{\hspace{2pt}}
C{1.2cm} C{1.2cm} @{\hspace{2pt}}
C{1.2cm} C{1.2cm} @{\hspace{2pt}}
C{1.2cm} C{1.2cm} @{\hspace{2pt}}
C{1.2cm} C{1.2cm} @{\hspace{2pt}}
C{1.2cm} C{1.2cm}
}
\hline\hline
\multicolumn{1}{c}{} &
\multicolumn{2}{c}{\parbox[c][\height][c]{2.5cm}{\centering $N_{\rm det(in)}=1$\tnote{a}}} &
\multicolumn{2}{c}{\parbox[c][\height][c]{2.5cm}{\centering $N_{\rm det(in)}=1,000$}} &
\multicolumn{2}{c}{\parbox[c][\height][c]{2.5cm}{\centering $N_{\rm det(in)}=5,000$}} &
\multicolumn{2}{c}{\parbox[c][\height][c]{2.5cm}{\centering $N_{\rm det(in)}=10,000$}} &
\multicolumn{2}{c}{\parbox[c][\height][c]{2.5cm}{\centering $N_{\rm det(in)}=20,000$}} \\
\cmidrule(lr){2-3} \cmidrule(lr){4-5} \cmidrule(lr){6-7}
\cmidrule(lr){8-9} \cmidrule(lr){10-11}
State &
\multicolumn{2}{c}{\%T=0.0\tnote{b}} &
\multicolumn{2}{c}{\%T=0.0--4.3\tnote{b}} &
\multicolumn{2}{c}{\%T=3.1--15.9\tnote{b}} &
\multicolumn{2}{c}{\%T=8.1--23.7\tnote{b}} &
\multicolumn{2}{c}{\%T=16.7--34.7\tnote{b}} \\
\cmidrule(lr){2-3} \cmidrule(lr){4-5} \cmidrule(lr){6-7}
\cmidrule(lr){8-9} \cmidrule(lr){10-11}
&
$P$\tnote{c} & ($P$;$Q$)\tnote{d} &
$P$\tnote{c} & ($P$;$Q$)\tnote{d} &
$P$\tnote{c} & ($P$;$Q$)\tnote{d} &
$P$\tnote{c} & ($P$;$Q$)\tnote{d} &
$P$\tnote{c} & ($P$;$Q$)\tnote{d} \\
\midrule
\AP{\rm X}{1} & 6.154  & 0.543 & 3.342  & 0.265 & 1.453 & 0.068 & 0.884 & 0.031 & 0.411 & 0.011 \\
\APP{1}{1}    & 7.666  & 1.517 & 2.941  & 0.853 & 1.240 & 0.660 & 0.805 & 0.458 & 0.449 & 0.249 \\
\AP{1}{3}     & 2.819  & 1.137 & 2.361  & 1.111 & 1.057 & 0.504 & 1.060 & 0.380 & 0.531 & 0.176 \\
\APP{1}{3}    & 7.693  & 1.080 & 3.164  & 0.908 & 1.318 & 0.666 & 0.863 & 0.449 & 0.484 & 0.237 \\
\AP{1}{1}     & 10.103 & 1.544 & 9.599  & 0.728 & 7.050 & 0.514 & 5.184 & 0.381 & 3.460 & 0.169 \\
\AP{2}{3}     & 8.846  & 1.225 & 8.413  & 0.744 & 6.311 & 0.479 & 4.326 & 0.304 & 3.121 & 0.128 \\
\APP{2}{3}    & 9.258  & 5.033 & 8.440  & 2.549 & 4.174 & 0.695 & 3.092 & 0.376 & 1.574 & 0.199 \\
\AP{2}{1}     & 14.460 & 2.988 & 9.019  & 0.764 & 3.984 & 0.580 & 3.066 & 0.423 & 1.652 & 0.284 \\
\APP{2}{1}    & 2.337  & 1.750 & 5.108  & 1.652 & 3.626 & 0.586 & 3.096 & 0.315 & 1.631 & 0.174 \\
\APP{3}{3}    & 28.242 & 6.054 & 18.471 & 4.875 & 6.634 & 1.116 & 3.917 & 0.544 & 1.850 & 0.250 \\
\AP{3}{1}     & 13.467 & 2.329 & 11.648 & 2.715 & 7.022 & 2.388 & 5.452 & 2.003 & 4.087 & 1.672 \\
\AP{3}{3}     & 8.305  & 2.481 & 6.899  & 2.622 & 4.712 & 1.358 & 3.379 & 1.051 & 2.324 & 0.826 \\
\hline\hline
\end{tabular}
\begin{tablenotes}
\footnotesize
\item[a]{
The MUE values characterizing the \ccp\ and \eomccp\ calculations at $N_{\rm det(in)} = 1$ are identical
to those obtained with CCSD and EOMCCSD. The corresponding \ccpq\ MUEs are equivalent to those obtained
with CR-CC(2,3) (the ground state) and CR-EOMCC(2,3) (excited states).}
\item[b]{
The \%T for a given $N_{\rm det(in)}$ is the percentage ($N_{\rm det(in)} = 1$) or the range of percentages
($N_{\rm det(in)} = 1,000\mbox{--}20,000$) of the $S_{z}=0$ triply excited determinants
captured by the CIPSI runs for the lowest-energy states of the relevant symmetries
[${\rm X}\,^{1}{\rm A}^{\prime}(C_{s})$ for the ${^{1}}{\rm A}^{\prime}$ and ${^{3}}{\rm A}^{\prime}$
states and $1 \,^{1}{\rm A}^{\prime\prime}(C_{s})$ for the ${^{1}}{\rm A}^{\prime\prime}$ and
${^{3}}{\rm A}^{\prime\prime}$ states]
at the various geometries of \ce{H2O} used to construct the ground- and excited-state potentials 
considered in the present study.}
\item[c]{
The MUE values characterizing the ground-state \ccp\ and excited-state \eomccp\ potentials relative to their
CCSDT and EOMCCSDT parents.}
%
\item[d]{
The MUE values characterizing the ground- and excited-state \ccpq\ potentials relative to their
CCSDT and EOMCCSDT parents.}
%
\end{tablenotes}
\end{threeparttable}
\end{table*}

\newpage
\clearpage
\pagebreak
\begin{table*}[ht!]
\centering
\footnotesize
\begin{threeparttable}
\caption{The NPE values, in millihartree, relative to CCSDT/EOMCCSDT characterizing the
ground- and 11 excited-state potential cuts of the water molecule, as described by the TZ
basis set of Ref.\ \cite{nmapp3}, along the O--H bond-breaking coordinate corresponding
to the $\ce{H2O} \rightarrow {\rm H} + {\rm OH}$ dissociation obtained with the CIPSI-based
\ccp/\eomccp\ and \ccpq\ approaches examined in the present work. The $P$ spaces used in
the \ccp\ and \eomccp\ calculations were defined as all singly and doubly excited
determinants and subsets of triply excited determinants extracted from the terminal $|\Psi^{\rm(CIPSI)}\rangle$
wave functions obtained in the CIPSI runs for the lowest-energy states of the relevant symmetries,
as described in the main text (see, also, footnote `b'). The $Q$ spaces used to construct the \ccpq\
corrections $\delta_{\mu}(P;\!Q)$ consisted of the remaining triples not captured by CIPSI.}
\label{table5}
\renewcommand{\arraystretch}{1.5}
\setlength{\tabcolsep}{1.2pt}
\footnotesize
\begin{tabular}{
C{1.2cm} @{\hspace{2pt}}
C{1.2cm} C{1.2cm} @{\hspace{2pt}}
C{1.2cm} C{1.2cm} @{\hspace{2pt}}
C{1.2cm} C{1.2cm} @{\hspace{2pt}}
C{1.2cm} C{1.2cm} @{\hspace{2pt}}
C{1.2cm} C{1.2cm}
}
\hline\hline
\multicolumn{1}{c}{} &
\multicolumn{2}{c}{\parbox[c][\height][c]{2.5cm}{\centering $N_{\rm det(in)}=1$\tnote{a}}} &
\multicolumn{2}{c}{\parbox[c][\height][c]{2.5cm}{\centering $N_{\rm det(in)}=1,000$}} &
\multicolumn{2}{c}{\parbox[c][\height][c]{2.5cm}{\centering $N_{\rm det(in)}=5,000$}} &
\multicolumn{2}{c}{\parbox[c][\height][c]{2.5cm}{\centering $N_{\rm det(in)}=10,000$}} &
\multicolumn{2}{c}{\parbox[c][\height][c]{2.5cm}{\centering $N_{\rm det(in)}=20,000$}} \\
\cmidrule(lr){2-3} \cmidrule(lr){4-5} \cmidrule(lr){6-7}
\cmidrule(lr){8-9} \cmidrule(lr){10-11}
State &
\multicolumn{2}{c}{\%T=0.0\tnote{b}} &
\multicolumn{2}{c}{\%T=0.0--4.3\tnote{b}} &
\multicolumn{2}{c}{\%T=3.1--15.9\tnote{b}} &
\multicolumn{2}{c}{\%T=8.1--23.7\tnote{b}} &
\multicolumn{2}{c}{\%T=16.7--34.7\tnote{b}} \\
\cmidrule(lr){2-3} \cmidrule(lr){4-5} \cmidrule(lr){6-7}
\cmidrule(lr){8-9} \cmidrule(lr){10-11}
&
$P$\tnote{c} & ($P$;$Q$)\tnote{d} &
$P$\tnote{c} & ($P$;$Q$)\tnote{d} &
$P$\tnote{c} & ($P$;$Q$)\tnote{d} &
$P$\tnote{c} & ($P$;$Q$)\tnote{d} &
$P$\tnote{c} & ($P$;$Q$)\tnote{d} \\
\midrule
\AP{\rm X}{1} & 8.453  & 0.692  & 1.744  & 0.222  & 1.191  & 0.075 & 0.644  & 0.035 & 0.396 & 0.015 \\
\APP{1}{1}    & 16.093 & 4.592  & 4.532  & 0.556  & 1.967  & 0.322 & 1.705  & 0.241 & 1.478 & 0.214 \\
\AP{1}{3}     & 5.004  & 0.442  & 4.431  & 0.617  & 2.279  & 0.375 & 3.161  & 0.264 & 2.078 & 0.155 \\
\APP{1}{3}    & 17.445 & 3.303  & 4.967  & 0.532  & 2.377  & 0.336 & 2.135  & 0.221 & 1.792 & 0.224 \\
\AP{1}{1}     & 21.043 & 4.637  & 21.133 & 0.502  & 17.668 & 0.789 & 11.340 & 0.513 & 8.503 & 0.487 \\
\AP{2}{3}     & 20.127 & 3.712  & 19.157 & 0.530  & 17.008 & 0.678 & 9.436  & 0.463 & 8.207 & 0.399 \\
\APP{2}{3}    & 29.017 & 13.044 & 24.403 & 4.875  & 7.586  & 0.559 & 4.900  & 0.421 & 2.587 & 0.213 \\
\AP{2}{1}     & 33.260 & 8.330  & 33.344 & 2.574  & 10.982 & 0.659 & 9.262  & 0.536 & 4.990 & 0.643 \\
\APP{2}{1}    & 7.971  & 3.983  & 14.093 & 2.814  & 5.408  & 0.389 & 4.300  & 0.405 & 2.844 & 0.280 \\
\APP{3}{3}    & 64.664 & 41.943 & 47.469 & 14.940 & 9.412  & 1.474 & 6.643  & 0.541 & 2.744 & 0.262 \\
\AP{3}{1}     & 30.847 & 4.376  & 30.966 & 5.012  & 15.483 & 5.282 & 10.661 & 5.198 & 9.084 & 5.065 \\
\AP{3}{3}     & 21.562 & 5.670  & 16.356 & 6.617  & 14.339 & 2.605 & 8.617  & 2.461 & 7.195 & 2.411 \\
\hline\hline
\end{tabular}
\begin{tablenotes}
\footnotesize
\item[a]{
The NPE values characterizing the \ccp\ and \eomccp\ calculations at $N_{\rm det(in)} = 1$ are identical
to those obtained with CCSD and EOMCCSD. The corresponding \ccpq\ NPEs are equivalent to those obtained
with CR-CC(2,3) (the ground state) and CR-EOMCC(2,3) (excited states).}
\item[b]{
The \%T for a given $N_{\rm det(in)}$ is the percentage ($N_{\rm det(in)} = 1$) or the range of percentages
($N_{\rm det(in)} = 1,000\mbox{--}20,000$) of the $S_{z}=0$ triply excited determinants
captured by the CIPSI runs for the lowest-energy states of the relevant symmetries
[${\rm X}\,^{1}{\rm A}^{\prime}(C_{s})$ for the ${^{1}}{\rm A}^{\prime}$ and ${^{3}}{\rm A}^{\prime}$
states and $1 \,^{1}{\rm A}^{\prime\prime}(C_{s})$ for the ${^{1}}{\rm A}^{\prime\prime}$ and
${^{3}}{\rm A}^{\prime\prime}$ states]
at the various geometries of \ce{H2O} used to construct the ground- and excited-state potentials
considered in the present study.}
\item[c]{
The NPE values characterizing the ground-state \ccp\ and excited-state \eomccp\ potentials relative to their
CCSDT and EOMCCSDT parents.}
%
\item[d]{
The NPE values characterizing the ground- and excited-state \ccpq\ potentials relative to their
CCSDT and EOMCCSDT parents.}
%
\end{tablenotes}
\end{threeparttable}
\end{table*}

\end{document}